\documentclass[twocolumn,preprintnumbers,amsmath,amssymb]{revtex4}
\usepackage{graphicx}
\usepackage{dcolumn}
\usepackage{amsmath}
\usepackage[utf8]{inputenc}
\usepackage{natbib}
\usepackage{cases}
\usepackage{bm}
\usepackage{float} 
\usepackage{color}
\usepackage{soul}
\usepackage{ulem}
\definecolor{myhighlight}{RGB}{255,255,0} 
\sethlcolor{myhighlight}
\soulregister{\cite}{7}
\soulregister{\ref}{7}

\begin{document}
\preprint{}
\title{The controlled exciton transport of the Multi-chain system by cavity-dressed energy level crossings and anticrossings }

\author{Jia-Hui Wang}
\affiliation{College of Physics and Electronic Engineering, Northwest Normal University, Lanzhou 730070, China.}
\author{Yu-Ren Shi}
\affiliation{College of Physics and Electronic Engineering, Northwest Normal University, Lanzhou 730070, China.}
\author{Ji-Ming Gao}
\affiliation{College of Physics and Electronic Engineering, Northwest Normal University, Lanzhou 730070, China.}
\author{Zi-Fa Yu}
\affiliation{College of Physics and Electronic Engineering, Northwest Normal University, Lanzhou 730070, China.}
\author{Ju-Kui Xue}
\affiliation{College of Physics and Electronic Engineering, Northwest Normal University, Lanzhou 730070, China.}
\author{Fang-Qi Hu}
\thanks{Email: thufang2008@163.com}
\begin{abstract}
The performance of various quantum devices is fundamentally linked to the control of exciton transport. To explore this, we study the exciton transport of the two-dimensional multi-chain systems with different coupling configurations in an optical cavity. Two types of the chains—the homogeneous and heterogeneous coupling chain, as well as two inter-chain coupling conformations—the square and triangular arrangements, are considered. The effects of the inter-chain coupling, the dimerization parameter, the cavity, the length and number of the chains on exciton transport are systematically investigated for different coupling configurations through the spectra, the Hopfield coefficients, and the steady-state dynamics of the system. The results show that in the absence of a cavity the exciton transport currents and efficiency are determined by the exciton distribution across the multi-chain system. However, when a cavity is introduced the exciton transport can be significantly enhanced or suppressed by the polariton formation at the cavity-dressed energy level crossings and anticrossings near zero-energy modes, where the coherent excitation and Landau-Zener transitions occur. Meanwhile, we discover that the discontinuous and extremal points in the second-order partial derivatives of the photon Hopfield coefficients with respect to the inter-chain coupling and the dimerization parameter correspond respectively to the crossings and anticrossings at the extreme points of the photon occupation number. Additionally, exciton transport efficiency is closely related to the odevity of both chain length and chain number, and exhibits oscillatory behaviour. This work provides critical insights into the exciton transport mechanism in multi-chain–cavity system and theoretical basis for designing high-performance excitonic devices with tunable transport properties.  

\end{abstract}


\maketitle


\section{Introduction}
In quantum device, energy and information are primarily transported by three carriers—electron, photon, and exciton. Optimizing and controlling the efficiency, the speed, the distance, and the current of carrier transport are the center to realize the functions of various quantum devices, such as organic solar cells \cite{wurfel2015impact,lu2022simultaneously,melianas2017charge}, organic light-emitting diodes \cite{shan2011transport,tanase2003unification,shen2022through}, quantum dot solids \cite{zhang2022ultrafast}, organic semiconductor devices \cite{sneyd2021efficient,hagenmuller2017cavity}, quantum simulation, and quantum computation \cite{georgescu2014quantum,yang2019quantum} etc. Exciton transport is influenced by multiple factors, including the dissipation \cite{cainelli2021exciton,zhu2024quantum,bose2022effect}, exciton-exciton annihilation \cite{wittmann2020energy}, disorder \cite{zhu2022single,quenzel2022plasmon,schachenmayer2015cavity,allard2022disorder,balasubrahmaniyam2023enhanced,ribeiro2022multimode}, defect sites \cite{wagner2011bound,akselrod2014visualization}, and topological property \cite{buendia2024long,allard2023multiple,nie2021topology}. Consequently, extensive research focuses on revealing the exciton transport mechanisms and properties in diverse quantum systems \cite{engel2007evidence,menke2013tailored}, and improving transport properties by reducing and avoiding detrimental factors. 

In experiment, the one-dimensional (1D) system carrying electron, photon, or exciton transport have been realized through various platforms, including organic aggregates molecule chains\cite{su1979solitons}, superconducting qubit circuits \cite{roushan2017spectroscopic,xu2018emulating,blais2003tunable}, Rydberg-atom chains \cite{barredo2015coherent,orioli2018relaxation}, plasmons \cite{su2021perovskite,seitz2020exciton}, semiconductors\cite{sneyd2021efficient,giannini2022exciton,muller2023directed}. At room temperature the strong exciton-photon coupling between organic aggregates and microcavities \cite{lidzey1998strong,lidzey1999room,kena2008strong,tischler2005strong,coles2014polariton} or plasmons \cite{bellessa2014strong,bellessa2004strong,zengin2015realizing,li2018plasmon} can result in new applications like polariton lasing \cite{kena2010room}, in which the polaritons form \cite{kavokin2003cavity,agranovich2003cavity}, and the molecular energy structures are modified \cite{galego2015cavity}. The polariton formation significantly modulates exciton transport properties containing the efficiency \cite{wei2019enhanced,wei2022cavity,feist2015extraordinary}, rate \cite{wang2021polariton,aroeira2023theoretical}, and distance \cite{du2018theory,myers2018polariton,sandik2024cavity,takazawa2010fraction} because the decay of excitons on chains is suppressed.

The hybrid system composed of a 1D molecular chain coupled to an optical cavity is well-described by the Tavis-Cummings model \cite{tavis1968exact}. The chain with the staggered hopping interaction is depicted by the Su-Schrieffer-Heeger (SSH) model \cite{su1979solitons}, which exhibits topologically trivial and nontrivial phases for even-numbered the two-level systems (TLSs) \cite{cai2019observation}. In the topologically nontrivial phase, even-length chains host a small gap bewteen two localized nearly-zero-energy modes within the topological bandgap, enabling hybrid edge modes formation from the left and the right edge states. These topologically protected edge states enhance exciton transport in finite SSH chains, as confirmed theoretically and experimentally \cite{chang2020probing,wei2022cavity}. The gap’s origin lies in energy-level anticrossings, where the Landau–Zener (LZ) transitions occur. In the two-dimensional (2D) and three-dimensional tungsten disulfide material, the anti-crossing dispersions were presented, in which the enhancement of exciton transport capability was corroborated in experiment \cite{guo2022boosting,yuan2017exciton}. Theoretical study have explored cavity-coupled system with multichains \cite{kvande2023finite}, suggesting that the exciton transport properties of the higher-dimensional ($\geq1$) systems warrant further investigation.

Motivated by the recent studies on the high dimensional tungsten disulfide, in this paper we devote to the studies of the inter-chain joint effect on exciton transport in 2D TLSs array within a cavity. Each chain contains two alternative coupling types, i.e., the homogeneous (HO) and the heterogeneous (HE) coupling, and two conformations of the inter-chain coupling are set, i.e., square and triangle types. Firstly, the energy spectra of the strong exciton-cavity coupled system are obtained by diagonalizing numerically the Hamiltonian for six configurations, analyzing their dependence on inter-chain coupling strength and dimerization parameter. Meanwhile, we calculate the von Neumann entropy of exciton to confirm the locations of the crossings and anticrossings, and the second order partial derivative of the photon Hopfield coefficients to show the change of the occupation number of the cavity photon. What's more, the effects of the inter-chain coupling, the dimerization parameter, the cavity, the TLS transition frequency detuning, the number and length of the chains on the steady-state dynamics of exciton transport are investigated for different configurations by numerically solving the Lindblad quantum master equation under the TLS and TLS+cavity drive. In the absence of cavity, the channel of exciton transport only is the multichains, in which its current and efficiency depend on the distribution of exciton on the whole multichains. Importantly, the exciton transport in the hybrid system correlates strongly with the crossings/anticrossings of dressed near-zero-energy mode, where the coherent excitation and LZ transition mediate the excitation exchange between the multichains and the cavity. Notably, polaritons play pivotal roles in controlling exciton transport through the crossings/anticrossings. In addition, the second order partial derivatives of the photon Hopfield coefficients serve as reliable metrics for identifying the extremal points in photon occupation number and the precise locations of the crossings/anticrossings. Further, we present the optimal parameters regions with the TLS transition frequency detuning, the inter-chain coupling strength, and the dimerization parameter. Finally, the exciton transport dynamics exhibit the odevity-dependent oscillations tied to the number and length of the chains.

The remainder of this paper is organised as follows: Sec. \ref{sec2} presents the theoretical framework, including the Hamiltonian of the multichain-cavity coupling system (in Sec. \ref{Hamiltonian}), Von Neumann entropy formulation for excitons (in Sec. \ref{von Neumann entropy}), the Lindblad quantum master equation of the system with an external driving field (in Sec. \ref{Transport dynamics}), and the steady-state exciton transport currents and efficiency (in Sec. \ref{Efficiency of exciton transport}). In Sec. \ref{sec3} the energy spectra are given by numerically diagonalizing the Hamiltonian. In Sec. \ref{sec4} we simulate the steady-state exciton dynamics across multi-chain configurations, analyzing the effects of the cavity, the inter-chain coupling, the dimerization parameter, the TLS transition frequency detuning (in Sec. \ref{subsec4.1}), the number and length of the chains (in Sec. \ref{subsec4.2}) on exciton transport. Section \ref{sec5} is a brief summary and future perspective.
\begin{figure}[ht]
\centering
\includegraphics[width=8.5cm]{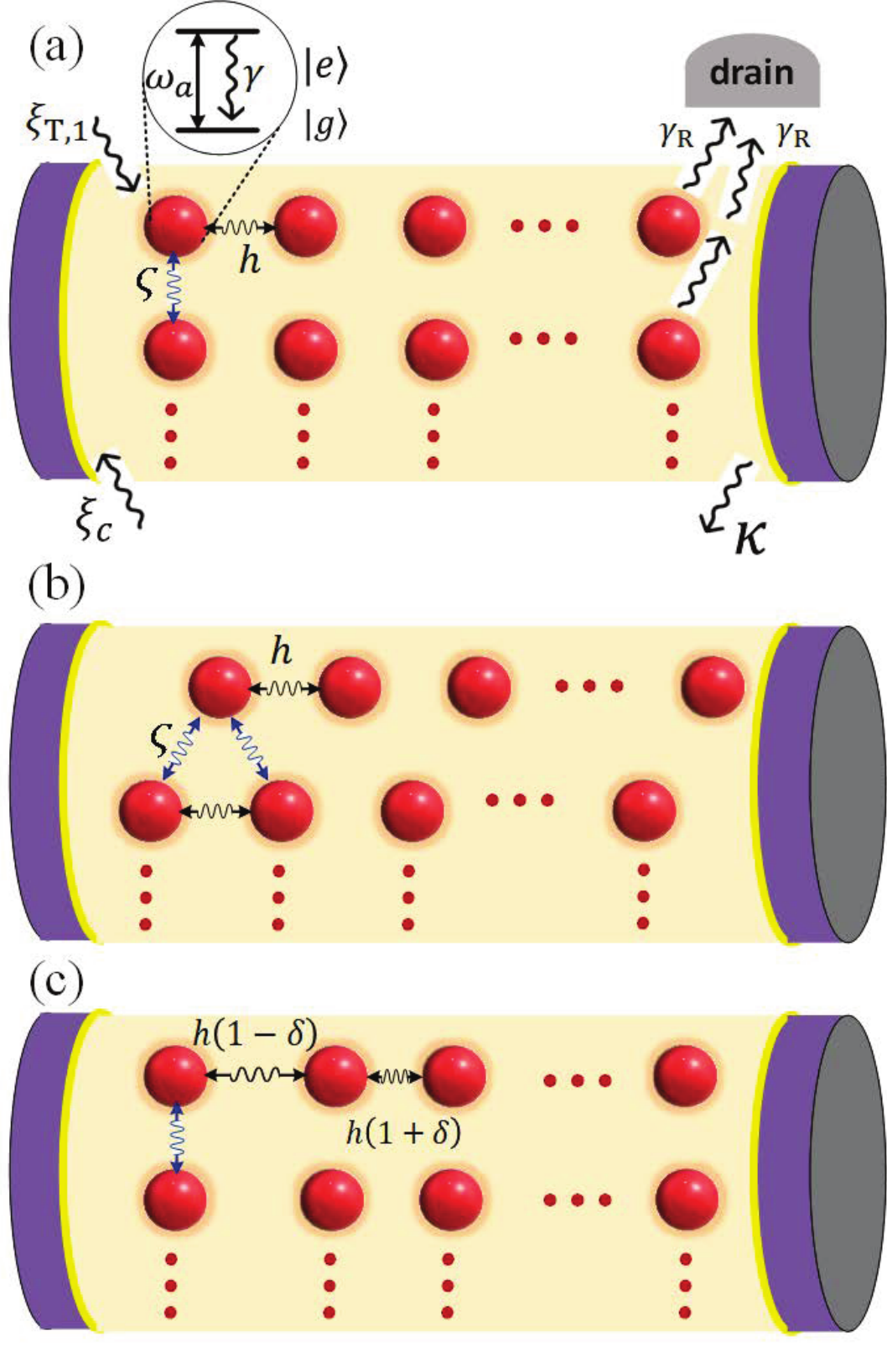}
\caption{\label{FIG:scheme graph01} (color online) The schematic representation of the hybrid system consisting of the 2D TLSs array within a cavity for three configurations HO ST (a), HO TT (b), and HE ST (c). Red spheres represent the TLSs with ground state $|g\rangle$ and excited state $|e\rangle$ between which the transition frequency is $\omega_{\rm a}$ and the decay rate is $\gamma$. The decay rate of the imperfect cavity is $\kappa$. The 2D TLSs array can be driven by an external coherent field with driving frequency $\omega$ via the first TLS of the first row with driving amplitude $\xi_{{\rm T},1}$ or/and via the cavity with driving amplitude $\xi_{c}$. The excitons in the multichains finally outflow from the last column into a reservoir (drain) with rate $\gamma_{\rm R}$. $h$ is the intra-chain hopping strength between adjacent TLSs. $\delta$ is the dimerization strength of HE chain. $\zeta$ is the inter-chains couplings strength between the nearest TLSs.}
\end{figure}
\section{Model}\label{sec2}
\subsection{Hamiltonian}\label{Hamiltonian}
The considered system is a 2D molecular aggregate embedded in a single-mode cavity, which can be modelled as multiple parallel chains with number $L$ ($L\geqslant2$) as illustrated in Fig. 1. The $k$th chain comprises $N_{k}$ identical TLSs. For simplicity, we assume that all chains contain the same number of the TLSs ($N_{1}=N_{2}=...=N_{L}=N$). The Hamiltonian of the multichain-cavity coupled system can be expressed as
\begin{eqnarray}
H_{0}=H_{\rm e}+H_{\rm c}+H_{\rm ec}\label{eq:H0}
\end{eqnarray}
The $H_{\rm e}$ represents the Hamiltonian of the multichains, and contains two terms as
\begin{eqnarray}
H_{\rm e}=\sum^{L}_{k=1}H_{{\rm e},k}+H_{\rm e,inter-chain},\label{eq:He}
\end{eqnarray}
where $H_{{\rm e},k}$ are the term of the $k$th bare chain.

Considering the staggered hopping interaction between the nearest-neighbor TLS monomers in a chain, under the rotating frame of the driving frequency $\omega$ the Hamiltonian of the $k$th chain is described by the SSH model
\begin{align}
H_{{\rm e},k}=&-\sum_{i=1}^{N}\Delta_{\rm a}\sigma_{k,i}^{+}\sigma_{k,i}^{-}-\nonumber \\
&\sum_{i=1}^{N-1}h\left[1+(-1)^{i}\delta\right](\sigma_{k,i}^{+}\sigma_{k,i+1}^{-}+{\rm H.c.}),\label{eq:Hek SSH}
\end{align}
where $h$ characterizes the hopping amplitude between two nearest-neighbor TLS monomers in a chain, $\delta$ represents the dimerization strength, $\sigma_{k,i}^{+}=|e_{k,i}\rangle\langle g_{k,i}|$ ($\sigma_{k,i}^{-}=|g_{k,i}\rangle\langle e_{k,i}|$) is the raising (lowering) operator of the TLS with same energy $\omega_{\rm a}$ (setting $\hbar=1$) at the $i$th site of the $k$th chain with the ground state $|g_{k,i}\rangle$ and the excited state $|e_{k,i}\rangle$ because of the identical particle, $\Delta_{\rm a}=\omega-\omega_{\rm a}$ is the TLSs transition frequency detuning relative to driving field frequency $\omega$, and ${\rm H.c.}$ denotes the Hermitian conjugate. The SSH chain behaves entirely different property for $\delta=0$ and $\delta\neq0$ corresponding to homogeneous and heterogeneous coupling in a molecular chain respectively, because the topologically nontrivial regime exists for the latter. In order to emphasize the difference in exciton transport between the two coupling cases, we designate separately them as the homogeneous (HO) and the heterogeneous (HE) coupling chain in the following content.

The inter-chain coupling term is denoted by the $H_{\rm e,inter-chain}$, whose form depends on the inter-chain coupling configuration in the 2D multichains. In this paper, we consider the two simplest configurations, the square type (ST) [Fig. 1(a)] and the triangle type (TT) [Fig. 1(b,c)]. The corresponding Hamiltonian can be respectively written as
\begin{widetext}
\begin{subequations}\label{eq:He inter-chain}
\begin{numcases}{H_{\rm e,inter-chain}=}
\zeta\sum^{L-1}_{k=1}\sum_{i=1}^{N}(\sigma_{k,i}^{+}\sigma_{k+1,i}^{-}+{\rm H.c.}) & \mbox{for ST}, \label{eq:H-inter-chain ST} \\
\zeta\sum^{P}_{k=1}\sum_{i=1}^{N-1}[\sigma_{2k-1,i}^{+}(\sigma_{2k,i}^{-}+\sigma_{2k,i+1}^{-})+\sigma_{2k-1,N}^{+}\sigma_{2k,N}^{-}+{\rm H.c.}]+\nonumber \\
\zeta\sum^{Q}_{k=1}\sum_{i=2}^{N}[\sigma_{2k,1}^{+}\sigma_{2k+1,1}^{-}+\sigma_{2k,i}^{+}(\sigma_{2k+1,i-1}^{-}+\sigma_{2k+1,i}^{-})+{\rm H.c.}] & \mbox{for TT}, \label{eq:H-inter-chain TT}
\end{numcases}
\end{subequations}
\end{widetext}
where for TT the upper bounds of $L$ in odd and even term are respectively
\begin{subequations}\label{eq:upper bound of L}
\begin{numcases}{P=}
(L-1)/2 & \mbox{$L$ is odd}, \\
L/2 & \mbox{$L$ is even},
\end{numcases}
\end{subequations}
\begin{subequations}\label{eq:upper bound of L}
\begin{numcases}{Q=}
(L-1)/2 & \mbox{$L$ is odd}, \\
(L-2)/2 & \mbox{$L$ is even},
\end{numcases}
\end{subequations}
and $\zeta$ denotes the inter-chain coupling strength, namely the hopping amplitude between two nearest-neighbor TLSs in two nearest-neighbor chains. Therefore, the multi-chain system exhibits six possible configurations based on the permutation and combination of the inter-chain and the intra-chain coupling types, which are expressed as HO ST, HO TT, HE ST, HE TT, HOCC-HE ST, and HO-HE TT. The HO ST, HO TT, and HE ST are schematically depicted in Fig. 1(a), Fig. 1(b), and Fig. 1(c) respectively. The HE TT configuration is obtained by replacing the HO chains in HO-TT with HE chains. For the hybrid cases, the HO and HE chains are arranged in a staggered pattern in both HO-HE ST and HO-HE TT, in which the odd-numbered and even-numbered chains are respectively HO and HE type.

In the rotating frame of the driving frequency $\omega$, the cavity radiation field is described by the Hamiltonian
\begin{eqnarray}
H_{\rm c}=-\Delta_{\rm c}c^{\dagger}c,\label{eq:Hc}
\end{eqnarray}
where $c^{\dagger}=|c\rangle\langle 0|$ ($c=|0\rangle\langle c|$) represents the photon creation (annihilation) operator, $\omega_{\rm c}$ is the cavity-mode frequency, $\Delta_{\rm c}=\omega-\omega_{\rm c}$ is the detuning of $\omega_{\rm c}$ with respect to $\omega$, $|c\rangle$ denotes the single-photon excited state of the cavity field, and $|0\rangle$ represents the joint vacuum state of the exciton-cavity system. To ensure the resonant condition between the cavity and the TLSs, we set $\omega_{\rm c}=\omega_{\rm a}$, hence $\Delta_{\rm a}=\Delta_{\rm c}=\Delta$.

In the rotating frame of $\omega$, the light-matter coupling between the multi-chain aggregate and the cavity is given by
\begin{eqnarray}
H_{\rm ec}=\sum^{L}_{k=1}\sum_{i=1}^{N}g_{k,i}(\sigma_{k,i}^{+}c+{\rm H.c.}),\label{eq:Hec}
\end{eqnarray}
where $g_{k,i}$ represents the coupling strength between the TLS at the $i$th site in the $k$th chain and the cavity field. For the sake of simplicity, we assume uniform coupling $g_{k,i}=g$ for all TLSs. Generally, $g$ depends on the collective Rabi splitting $\Omega_{\rm R}=2\sqrt{\sum_{k=1}^{L}\sum_{i=1}^{N}g^{2}_{k,i}}=2g\sqrt{LN}$.

In order to investigate the excitation dynamics, we introduce an external coherent drive in the exciton-cavity system. We consider two driving means, i.e., the driving field is exerted on the first site of the $k$th chain, as well as simultaneously on both this site and cavity, which can be described respectively by
\begin{subequations}\label{eq:Hd 1}
\begin{numcases}{H_{\rm d}=}
H_{\rm dT}=\xi_{{\rm T},k}(\sigma_{k,1}^{-}e^{i\omega t}+\sigma_{k,1}^{+}e^{-i\omega t})\label{eq:Hda 1}, \\
H_{\rm dTc}=H_{\rm dT}+\xi_{\rm c}(ce^{i\omega t}+c^{\dagger}e^{-i\omega t}) \label{eq:Hdc 1},
\end{numcases}
\end{subequations}
where $\omega$ is the driving frequency, $\xi_{{\rm T},k}$ ($\xi_{\rm c}$) denotes the driving strength for the TLS (cavity). The second mean is denoted by "TLS+cavity" in the following. Here, we set $k=1$, namely, only the first site in the first chain is driven. The two driving means were used to study the two-photon blockade in the cavity QED system \cite{hamsen2017two}. In the rotating frame of $\omega$, the Eq. (\ref{eq:Hd 1}) can be written as
\begin{subequations}\label{eq:Hd 2}
\begin{numcases}{H_{\rm d}=}
H_{\rm dT}=\xi_{{\rm T},1}(\sigma_{1,1}^{-}+\sigma_{1,1}^{+}) \label{eq:Hdc 2}, \\
H_{\rm dTc}=H_{\rm dT}+\xi_{c}(c+c^{\dagger})\label{eq:Hda 2}.
\end{numcases}
\end{subequations}

When the driving field is considered, the total Hamiltonian of the open quantum system is
\begin{eqnarray}
H=H_{0}+H_{\rm d}.\label{eq:H}
\end{eqnarray}
In the following, we assumpt that the driving field is weak so that the representation of the total Hamitonian $H$ is restricted to the zero- and single-excitation subspace with $\sum_{k=1}^{L}\sum_{i=1}^{N}\sigma_{k,i}^{+}\sigma_{k,i}^{-}+c^{\dagger}c=0$ and 1. The corresponding Hilbert subspace is spanned by the basis vectors $\{|0\rangle,|k,i\rangle,|c\rangle\}$, where $|k,i\rangle$ $(k=1,2,...,L$ and $i=1,2,...,N)$ is the excited state of the $i$th site in the $k$th chain.
\subsection{von Neumann entropy}\label{von Neumann entropy}
In order to find and distinguish accurately the locations of the energy level crossings and anticrossings of Hamiltonian (\ref{eq:H0}) in the parameter space of the inter-chain coupling strength $\zeta$ and the dimerization parameter $\delta$, one can calculate the von Neumann entropy of the excitons by 
\begin{eqnarray}
S(\rho_{\rm e})=-{\rm tr}(\rho_{\rm e}{\rm log}_{2}\rho_{\rm e}).\label{eq:entropy}
\end{eqnarray}
For the exciton-photon hybrid system, its eigenstates can be indicated as $|\psi_{\rm e-p}\rangle$. Thus the reduced density matrix $\rho_{\rm e}$ of the excitons is obtained by tracing out the degrees of freedom of the photons, i.e., $\rho_{\rm e}={\rm tr}_{\rm p}(|\psi_{\rm e-p}\rangle\langle\psi_{\rm e-p}|)$. According to Ref. \cite{hu2018effect}, the crossings (anticrossings) correspond to the saltatorial intersections (maxima or acclive intersections) of the von Neumann entropy with parameter. Therefore, the entanglement between upper and lower energy levels is maximal at the anticrossings, where the LZ transitions occur. The transition from anticrossing to crossing means the energy degeneracy \cite{hu2018effect}. At the crossings, the excitations are exchanged coherently between multichains and cavity. As a result, the boosted exciton transport in degenerate states was demonstrated \cite{tagarelli2023electrical,lin2012efficient}.
\subsection{Transport dynamics}\label{Transport dynamics}
For the system that exists dissipation at each TLS in the multichains, the driving field, and the coupling with environment (reservoir), its dynamics are governed by the Lindblad quantum master equation under the Born-Markov approximations and the rotating-wave approximation \cite{lindblad1976generators}. Then, the exciton transport dynamics for the Hamiltonian (\ref{eq:H}) can be investigated by the master equation
\begin{eqnarray}
\dot{\rho}=i[\rho,H]+\sum^{L}_{k=1}(D_{{\rm d},k}[\rho(t)]+D_{{\rm R},k}[\rho(t)]),\label{eq:Lindblad master equation}
\end{eqnarray}
where the first term in the righthand side accounts for the coherent evolution under the total Hamiltonian (\ref{eq:H}), the second term characterizes the decay of excitons due to the spontaneous emission with
\begin{eqnarray}
D_{{\rm d},k}[\rho]=\sum^{N}_{i=1}\gamma_{d}\mathcal{L}_{\sigma^{-}_{k,i}}[\rho]+\kappa\mathcal{L}_{\rm c}[\rho],\label{eq:exciton decay}
\end{eqnarray}
where $\gamma_{d}$ is the spontaneous emission rate of the TLSs, $\kappa$ is the decay rates of the cavity photons, and $\mathcal{L}_{x}[\rho]=x\rho x^{\dagger}-\dfrac{1}{2}\left\{\rho,x^{\dagger}x \right\}$ is the Lindblad superoperators with standard form. The term $D_{{\rm R},k}[\rho(t)]$ in Eq. (\ref{eq:Lindblad master equation}) describes the exciton exchange between chain ends and exciton drain, which is given by
\begin{eqnarray}
D_{{\rm R},k}[\rho]=\gamma_{\rm R}\bar{n}_{\rm R}\mathcal{L}_{\sigma_{k,N}^{+}}[\rho]+\gamma_{\rm R}(\bar{n}_{\rm R}+1)\mathcal{L}_{\sigma_{k,N}^{-}}[\rho],\label{eq:exciton exchange}
\end{eqnarray}
where $\gamma_{\rm R}$ is the exciton exchange rate, $\bar{n}_{\rm R}$ is the average exciton number in the drain. The last sites of each chain exist coupling with the drain, so $k\in\{1,2,...,L\}$. The two terms in Eq. (\ref{eq:exciton exchange}) represent respectively the inflow and outflow of excitons from the last column TLSs of the multichains to the drain. In order to show the net transportation of exciton, we set $\bar{n}_{\rm R}=0$, establishing unidirectional exciton transfer from the last column of the multichains to the drain without reflection, which is reasonable as the photosynthesis process in nature.

\subsection{Efficiency of exciton transport}\label{Efficiency of exciton transport}
\begin{figure*}[ht]
\centering
\includegraphics[width=15cm]{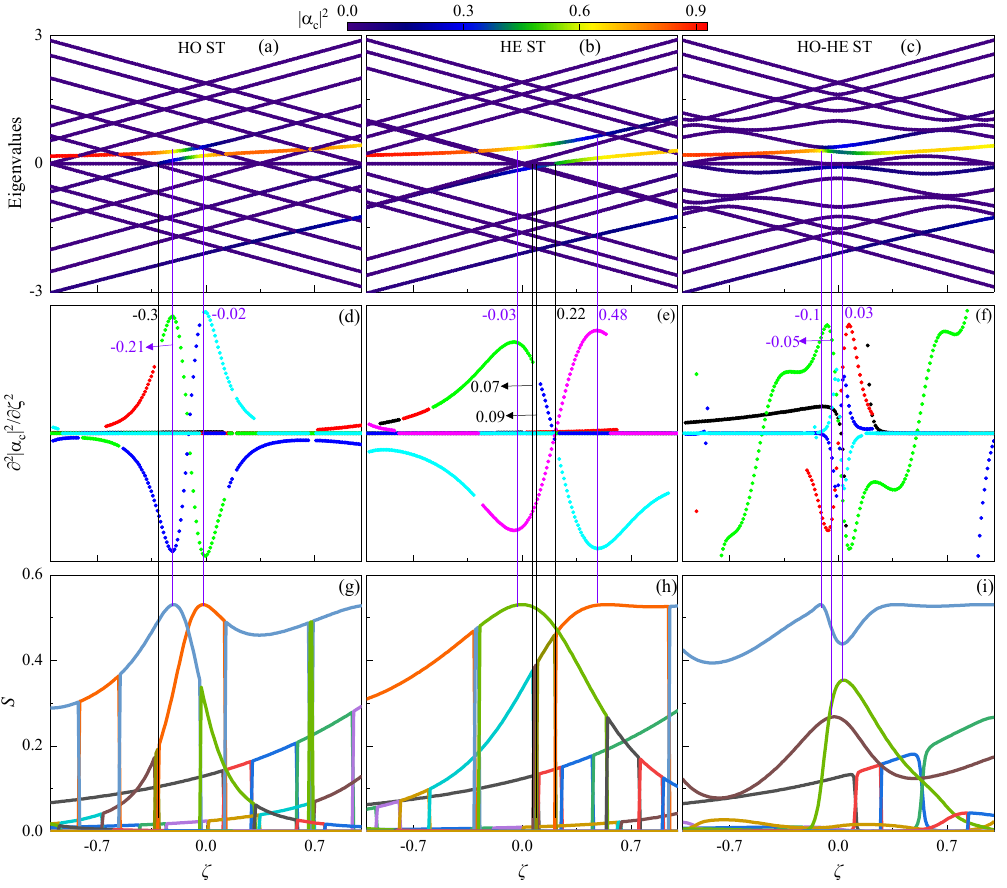}
\caption{\label{FIG:energy level zita-1} (color online) The energy spectra of the exciton-cavity coupling system as a function of the inter-chain coupling strength $\zeta$ for different configurations HO ST (a), HE ST (b), HO-HE ST (c), with the colormap representing the photon Hopfield coefficients by numerically diagonalizing Hamiltonian (\ref{eq:H0}) in single-excitation space. (d-f) The second order partial derivative of the photon Hopfield coefficients with respect to $\zeta$, which contains the locations of discontinuity and extremum corresponding to that existing evident changes in exciton transport dynamics. (g-i) The corresponding von Neumann entropy $S$ of excitons by calculating Eq. (\ref{eq:entropy}), in which the colors of the lines has not means just to distinguish each other. The black (violet) vertical lines and the data texts mark the locations of cavity-dressed energy level crossings (anticrossings) (a-c) producing obvious effect on exciton transport, the discontinuities (extrema) of the second order partial derivative of the photon Hopfield coefficients (d-f), as well as the saltatorial intersections (maxima or acclive intersections) of the von Neumann entropy (g-i). The corresponding parameters are $L=2$, $N=8$, $\Delta_{\rm a}=\Delta_{\rm c}=0$ eV, $h=1$ eV, $\delta=0.5$, and $\Omega_{R}=1$ eV.}
\end{figure*}
\begin{figure*}[ht]
\centering
\includegraphics[width=15cm]{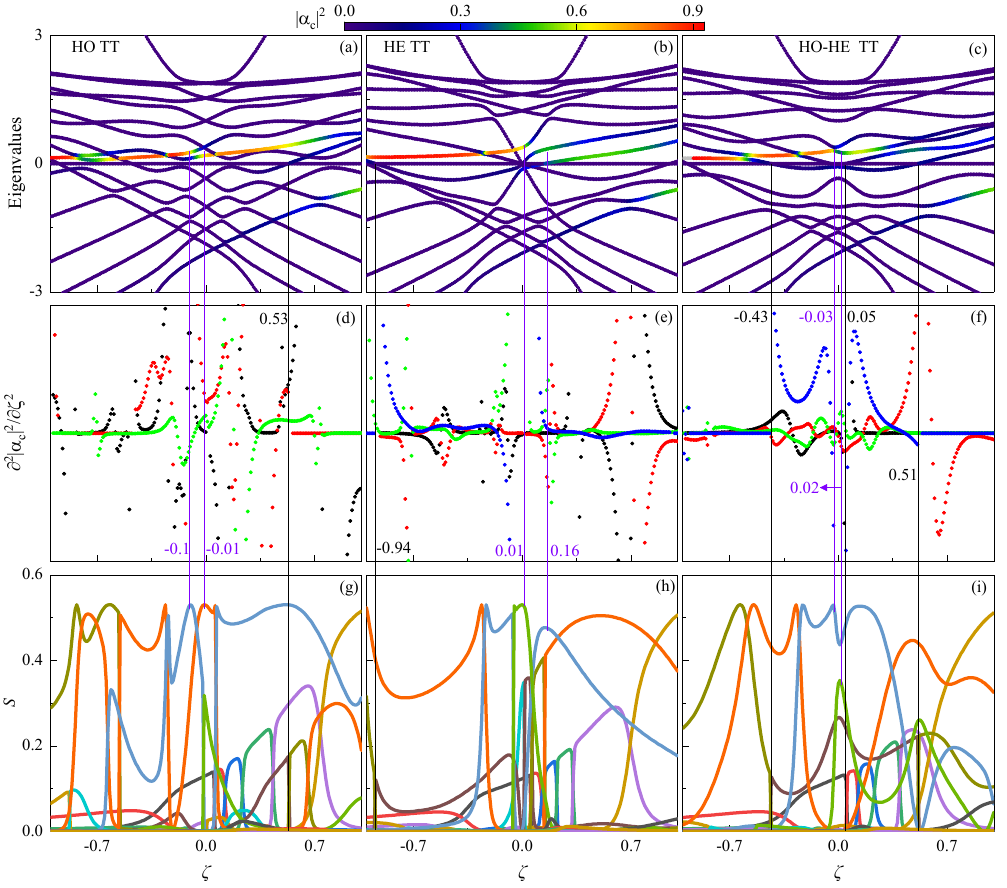}
\caption{\label{FIG:energy level zita-2} (color online) Similar as Fig \ref{FIG:energy level zita-1}, but for different configurations HO TT (a, d, and g), HE TT (b, e, and h), and HO-HE TT (c, f, and i).}
\end{figure*}
\begin{figure*}[ht]
\centering
\includegraphics[width=17cm]{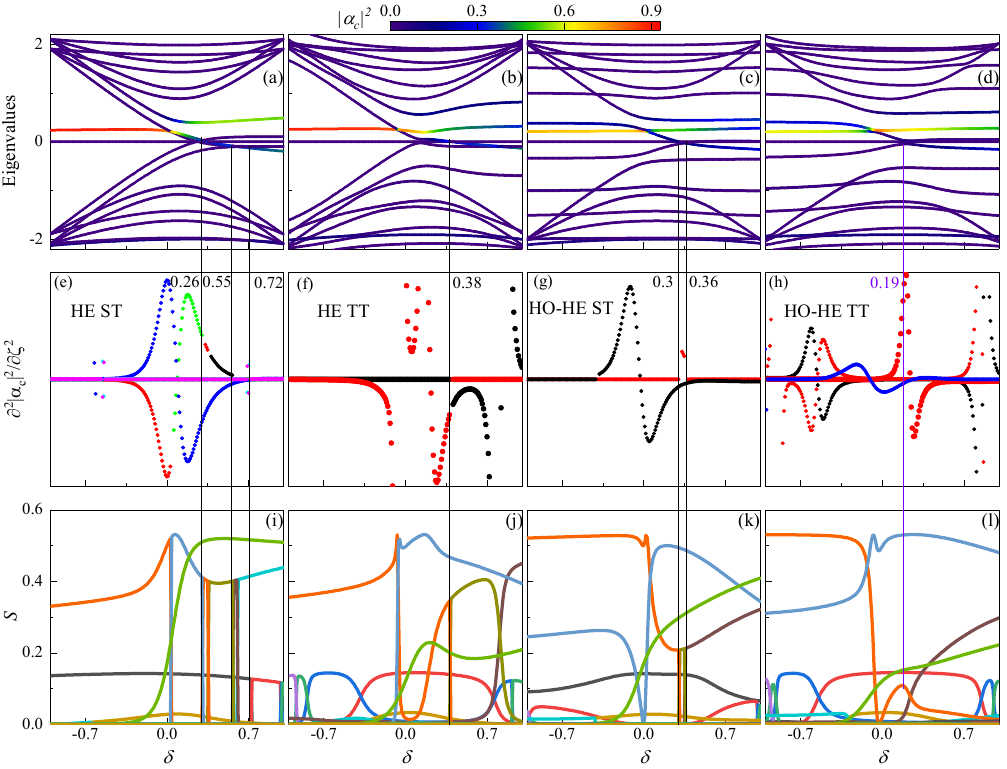}
\caption{\label{FIG:energy level delta-1} (color online) Similar as Fig \ref{FIG:energy level zita-1}, but for even chain $N=8$ and for different configurations HE ST (a, e, and i), HE TT (b, f, and j), HO-HE ST (c, g, and k), and HO-HE TT (d, h, and l) with dimerization parameter $\delta$. The parameter used is $\zeta=0.1$ eV, the others are the same as those in Fig. \ref{FIG:energy level zita-1}.}
\end{figure*}
\begin{figure*}[ht]
\centering
\includegraphics[width=17cm]{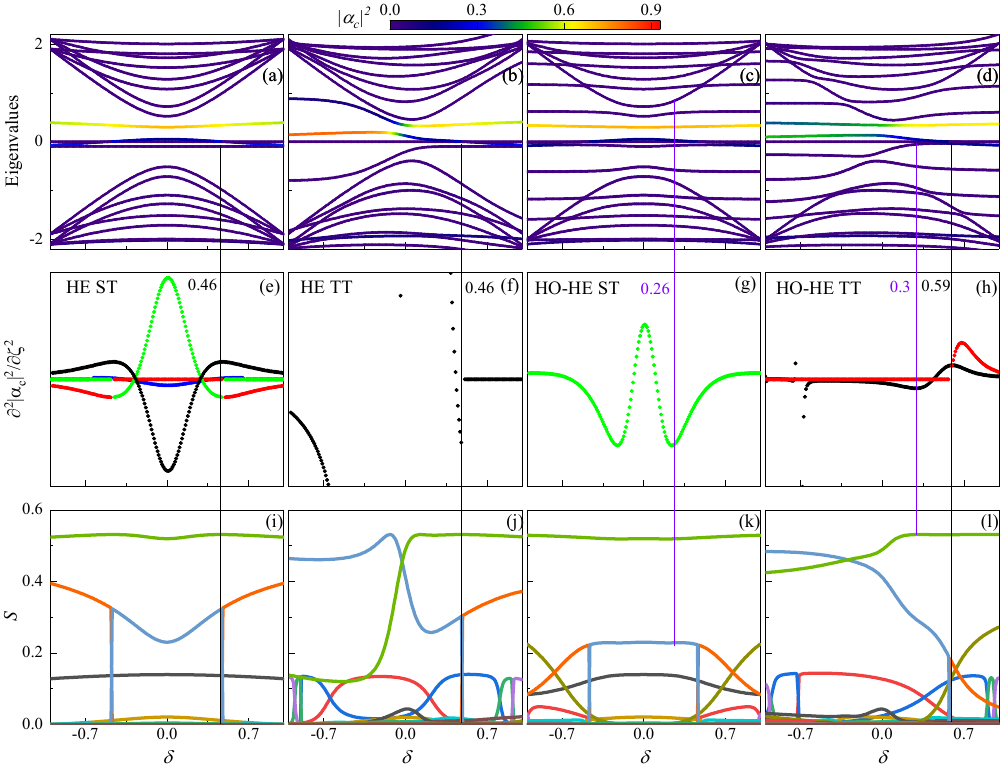}
\caption{\label{FIG:energy level delta-2} (color online) Similar as Fig \ref{FIG:energy level delta-1}, but for odd chain $N=9$.}
\end{figure*}

To quantify the exciton transport dynamics in the 2D multichain-cavity system, the exciton transport efficiency $\eta$ is defined and derived in this subsection. Generally, the exciton transport efficiency is characterized by the ratio of the outgoing $(I^{\rm s}_{\rm o})$ and input $(I^{\rm s}_{\rm i})$ exciton currents for steady state \cite{liu2019vibration,wei2019enhanced}, that is
\begin{eqnarray}
\eta=\lim_{t\to\infty}\frac{I_{\rm o}}{I_{\rm i}}=\frac{I_{\rm o}^{\rm s}}{I_{\rm i}^{\rm s}},\label{eq:efficiency}
\end{eqnarray}
where the superscript "s" refers to the steady state.
On the other hand, the exciton occupation number can reveal the transport mechanism and determine the currents. The exciton occupation number on the $j$th site of the $k$th chain is given by
\begin{eqnarray}
p_{k,j}(t)\equiv {\rm Tr}\{\rho\sigma^{+}_{k,j}\sigma^{-}_{k,j}\}.\label{eq:occupation}
\end{eqnarray}
Combining Eq. (\ref{eq:Lindblad master equation}) and Eq. (\ref{eq:occupation}), the time evolution of $p_{k,j}(t)$ obeys
\begin{align}
\dot{p}_{k,j}(t)=&i{\rm Tr}\{[\rho,H]\sigma^{+}_{k,j}\sigma^{-}_{k,j}\}+\nonumber\\
&{\rm Tr}\{\sum_{\mu={\rm d,R}}\sum^{L}_{k=1}D_{\mu,k}[\rho]\sigma^{+}_{k,j}\sigma^{-}_{k,j}\},\label{eq:occupation equation}
\end{align}
where the trace is operated throughout the Hilbert space of the total system. The lefthand side describes the rate of the change of $p_{k,j}(t)$ with time, thus the first term (the second term) of the righthand side represents the input (outgoing) exciton current at the $j$th site in the $k$th chain. According to the explanation, under the TLS and TLS+cavity drive the input exciton current of the multichains can be written respectively as
\begin{subequations}\label{eq:input exciton current}
\begin{numcases}{I_{\rm i}(t)=}
i{\rm Tr}\{[\rho,H_{\rm dT}]\sigma^{+}_{1,1}\sigma^{-}_{1,1}\}, \label{eq:input exciton current dt} \\
i{\rm Tr}\{[\rho,H_{\rm dT}]\sigma^{+}_{1,1}\sigma^{-}_{1,1}+[\rho,\xi_{c}(c+c^{\dagger})]c^{\dagger}c\}, \label{input exciton current dc}
\end{numcases}
\end{subequations}
which is related to the exciton injection at the first site in the $k$th chain or/and in cavity via the external coherent field per unit time. Substituting Eq. (\ref{eq:Hd 1}) for $H_{\rm dc}$ and $H_{\rm dTc}$ in Eq. (\ref{eq:input exciton current}), under the zero- and single- excitation subspace, the input exciton currents are simplified to
\begin{subequations}\label{eq:input exciton current 1}
\begin{numcases}{I_{\rm i}(t)=}
i\xi_{{\rm T},1}(\langle 1,1|\rho|0\rangle-{\rm H.c.}), \label{eq:input exciton current dt 1} \\
i\xi_{{\rm T},1}(\langle 1,1|\rho|0\rangle-{\rm H.c.})+i\xi_{\rm c}(\langle c|\rho|0\rangle-{\rm H.c.}). \label{input exciton current dc 1}
\end{numcases}
\end{subequations}
According to the Hermitian property of density matrix, under the TLS and TLS+cavity drive the input exciton currents are finally written as
\begin{subequations}\label{eq:input exciton current 2}
\begin{numcases}{I_{\rm i}(t)=}
2\xi_{{\rm T},1}{\rm Im}[\langle 0|\rho|1,1\rangle], \label{eq:input exciton current dt 2} \\
2\xi_{{\rm T},1}{\rm Im}[\langle 0|\rho|1,1\rangle]+2\xi_{c}{\rm Im}[\langle 0|\rho|c\rangle]. \label{input exciton current dc 2}
\end{numcases}
\end{subequations}

Analogous to the input exciton current, the outgoing exciton current can be defined as the outflow of exciton transfering from the last column in the multichains to the drain per unit time. According to the means of the third term of righthand in Eq. (\ref{eq:occupation equation}), it can be given by
\begin{eqnarray}
I_{\rm o}(t)={\rm Tr}[\sum^{L}_{k=1}D_{{\rm R},k}[\rho(t)]\sigma_{k,N}^{+}\sigma_{k,N}^{-}].\label{eq:outgoing exciton current 1}
\end{eqnarray}
Pluging the Eq. (\ref{eq:exciton exchange}) into the Eq. (\ref{eq:outgoing exciton current 1}), under the single-excitation assumption, the form of the outgoing exciton current is simplified as
\begin{eqnarray}
I_{\rm o}(t)=\gamma_{\rm R}\sum^{L}_{k=1}p_{k,N}, \label{eq:outgoing exciton current 2}
\end{eqnarray}
where $p_{k,N}$ is the occupation number at the last site of the chain $k$.

\section{Spectrum of coupling system}\label{sec3}
There are a large number of studies on the spectra of the SSH Hamiltonian for the open boundary conditions  \cite{sirker2014boundary,nie2020bandgap,wei2022cavity,cai2019observation}. The topological phases of the SSH chain depend on the odevity of TLS number in 1D chain and the qubit-qubit coupling configuration. According to the Ref. \cite{cai2019observation}, the dimerization parameter $\delta>0$ ($\delta<0$) corresponds to the topological winding number $\nu=1$ ($\nu=0$), i.e., the topologically nontrivial (trivial) regime. In Ref. \cite{sirker2014boundary}, the spectra show that there is single zero-energy mode in whole range of $\delta$ for odd-length chain and are two localized nearly-zero-energy modes in the topologically nontrivial regime $\delta>0$ for even-length chain. The finite gap between the two edge modes leads to the formation of the hybrid edge modes that overlap with each other, generating the polaritonic states when the edge modes are dressed by cavity due to the exciton-photon coupling. Comparing to the single chain system, the presence of the inter-chain coupling induces the spectral extension of the 2D multi-chain system, adding greatly to the complexity of the system dynamics.  In the exciton-cavity coupling system, the crossings and anticrossings of the edge modes dressed by the cavity can lead to the abundant exchange of excitation between the multichains and the cavity attributing to the coherent exchanges and the LZ transitions, where the polaritons increase or decrease. The polaritons act as intermediary in exciton transport, causing that the excitons are transported among site $(k,1)~ \rightarrow$ cavity $\rightarrow$ site $(k,N)$ via bypassing dissipative intermediate TLSs \cite{wei2022cavity}. Therefore, the exciton transport can be enhanced by the polaritons at the crossings and anticrossings.
\begin{figure}[htbp]
\centering
\includegraphics[width=8.5cm]{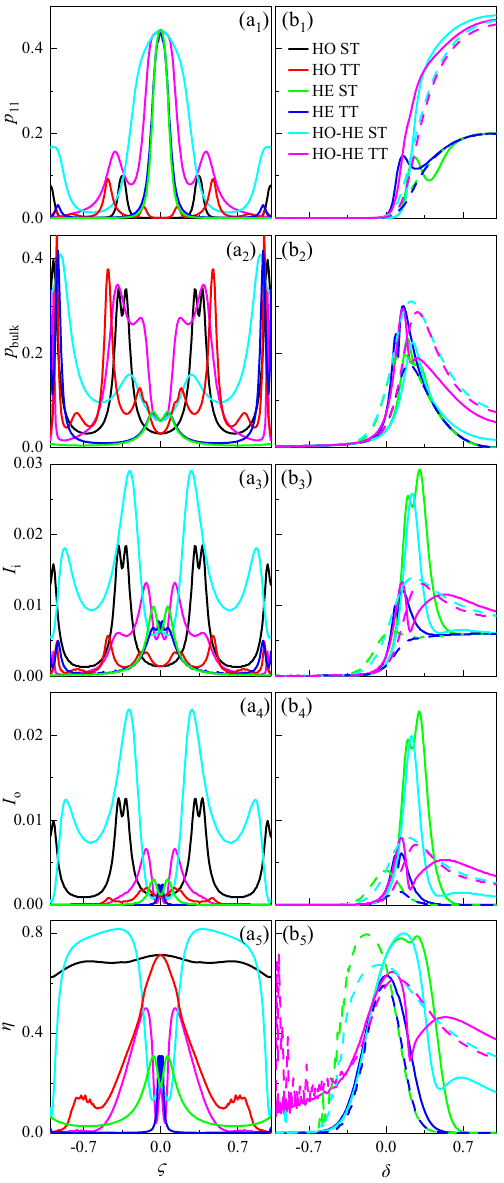}
\caption{\label{FIG:stady dynamic zita delta withoutC line} (color online) The dependence of the steady-state occupation numbers (a$_{1}$ and b$_{1}$) $p_{11}$ on site (1,1), (a$_{2}$ and b$_{2}$) $p_{\rm bulk}$ on bulk sites, (a$_{3}$ and b$_{3}$) input exciton current $I_{\rm i}$, (a$_{4}$ and b$_{4}$) outgoing exciton current $I_{\rm o}$, and (a$_{5}$ and b$_{5}$) transport efficiency $\eta$ on the inter-chain coupling strength $\zeta$ (a$_{1-5}$) and the dimerization parameter $\delta$ (b$_{1-5}$) for different configurations without cavity. In (b$_{1-5}$) the solid and dash lines correspond respectively to $N=8$ and 9. The parameters used are $\zeta=0.1$ eV, $\xi_{{\rm T},1}=\xi_{\rm c}=0.1$ eV, $\gamma_{\rm d}=0.01$ eV, $\gamma_{\rm R}=0.1$ eV, and the others are the same as those in Fig. \ref{FIG:energy level zita-1}.}
\end{figure}
\begin{figure*}[ht]
\centering
\includegraphics[width=17cm]{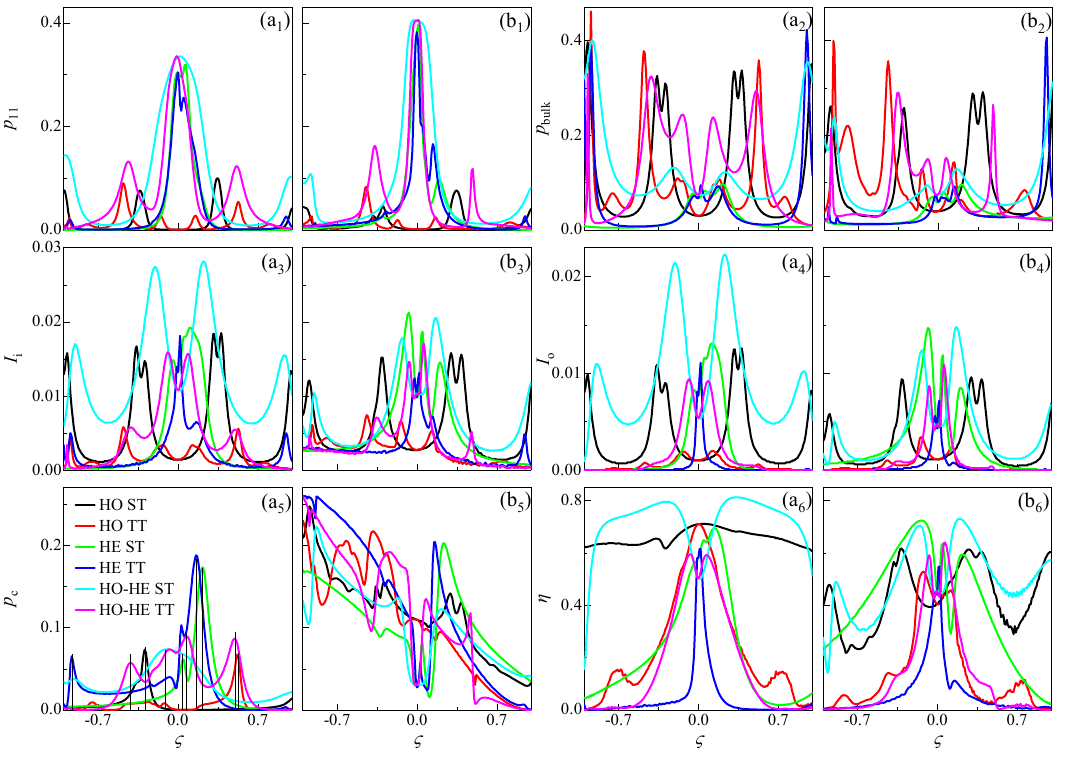}
\caption{\label{FIG:stady dynamic zita withC line} (color online) The dependence of the steady-state dynamics of the hybrid system on the inter-chain coupling strength $\zeta$ in the presence of cavity under TLS drive (a$_{1-6}$) and TLS+cavity drive (b$_{1-6}$). (a$_{5}$ and b$_{5}$) The occupation number of photon with $\zeta$, in which the extreme points are marked by the vertical fine lines, and the others are similar to Fig. \ref{FIG:stady dynamic zita delta withoutC line}. The parameter used is $\kappa=0.01$ eV, and the others are the same as those in Fig. \ref{FIG:stady dynamic zita delta withoutC line}.}
\end{figure*}
\begin{figure*}[ht]
\centering
\includegraphics[width=17cm]{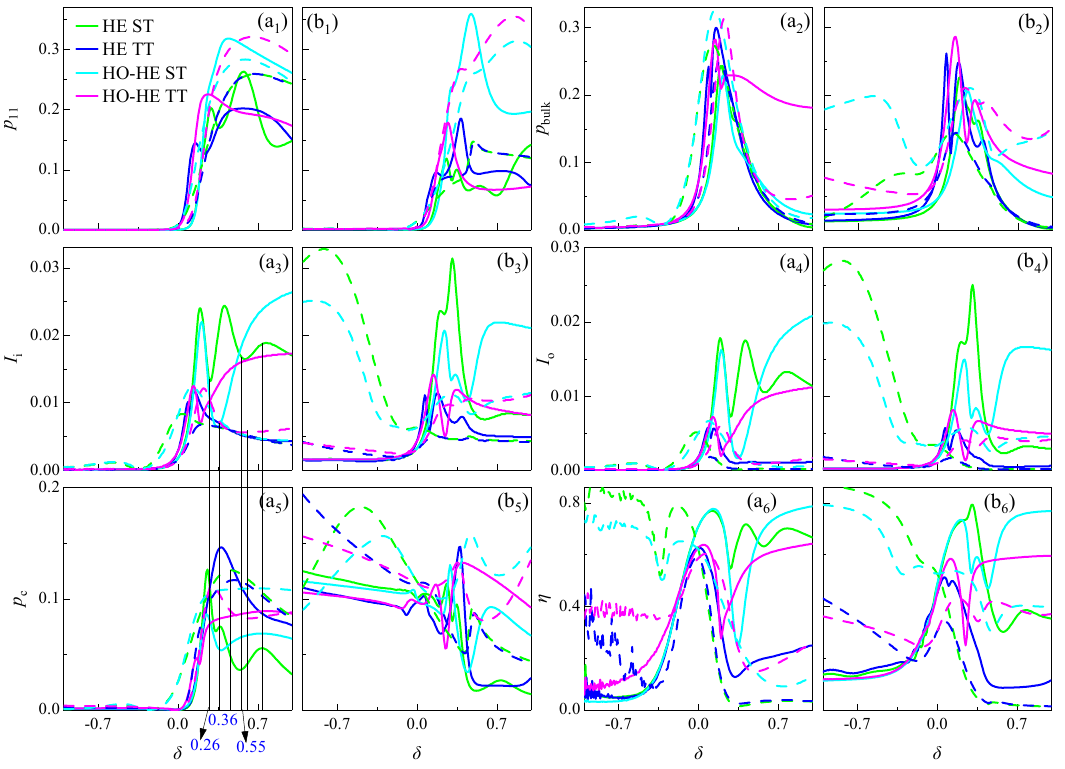}
\caption{\label{FIG:stady dynamic delta withC line} (color online) The dependence of the steady-state dynamics of the hybrid system on the dimerization parameter $\delta$, and the others are similar to Fig. \ref{FIG:stady dynamic zita withC line}. The parameters used are the same as those in Fig. \ref{FIG:stady dynamic zita withC line}.}
\end{figure*}

In order to analyse the spectral characteristics of the multichain-cavity coupled system, we numerically diagonalize the Hamiltonian (\ref{eq:H0}) to obtain its eigenvalues. The formation of polaritons is investigated through the photon Hopfield coefficients $|\alpha_{c}|^{2}$. When the $j$th normalized eigenvector of the Hamiltonian (\ref{eq:H0}) is expanded as $|\psi_{j}\rangle=\Sigma_{k}^{L}\Sigma_{i}^{N}\alpha_{k,i}|k,i\rangle+\alpha_{c}|c\rangle$, $|\alpha_{k,i}|^{2}$ and $|\alpha_{c}|^{2}$ denote the Hopfield coefficient of exciton and photon, which characterize the hybridization weights of exciton and photon components, respectively. The color maps Fig. \ref{FIG:energy level zita-1} (a-c), \ref{FIG:energy level zita-2} (a-c) and Fig. \ref{FIG:energy level delta-1} (a-d), \ref{FIG:energy level delta-2} (a-d) show respectively the spectra of the exciton-cavity coupled system as functions of the inter-chain coupling strength $\zeta$ and the dimerization parameter $\delta$ for different coupling configurations, and contain the photon Hopfield coefficient $|\alpha_{c}|^{2}$. One can see that the zero-energy modes exist in the spectra for all situations. $|\alpha_{c}|^{2}\neq0$ means that the energy level is dressed by the cavity mode. It is obvious that only the nearly-zero-energy modes are distinctly dressed by the cavity mode, generating corresponding polaritonic branches, while the high excited states are hardly dressed for the given exciton-cavity coupling strength. In addition, there are the crossings (anticrossings) in the spectra, in which the crossings (anticrossings) between the nearly-zero-energy modes dressed by the cavity mode are our focus. It can be seen that the photonic content distribution on the upper and lower polaritonic branches varies significantly with $\zeta$ and $\delta$. It means that the coherent exchange of excitation or the LZ transition of photons occurs, inducing the changes of the exciton transport efficiency and currents, which will be analysed in the next section by the steady-state dynamics. Further, it can be discovered that the number and locations of crossings (anticrossings) affecting transport depend on the coupling configurations of TLSs in the range of $\zeta$ and $\delta$. In comparison, the effect of $\zeta$ on the crossings (anticrossings) is more significant than that of $\delta$.

To systematically analyze the photon Hopfield coefficient $|\alpha_{c}|^{2}$, we plot the scatter maps of the second order partial derivatives $\partial^{2}|\alpha_{c}|^{2}/\partial\zeta^{2}$ in Fig. \ref{FIG:energy level zita-1} (d-f) and \ref{FIG:energy level zita-2} (d-f), as well as $\partial^{2}|\alpha_{c}|^{2}/\partial\delta^{2}$ in Fig. \ref{FIG:energy level delta-1} (e-i) and \ref{FIG:energy level delta-2} (e-i), which correlate with the obvious changes of the steady-state exciton transport dynamics. To determine and distinguish accurately the locations of the crossings and anticrossings in spectra, the von Neumann entropy $S$ of excitons is calculated through Eq. (\ref{eq:entropy}), and is shown in Fig. \ref{FIG:energy level zita-1} (g-i), \ref{FIG:energy level zita-2} (g-i), \ref{FIG:energy level delta-1} (i-l), and \ref{FIG:energy level delta-2} (i-l). The dressed nearly-zero-energy crossings (anticrossings) identified by $S$ are marked in the spectra, which correspond to the saltatorial intersections (maxima or acclive intersections) of $S$. Significantly, we discover that the discontinuous locations of $\partial^{2}|\alpha_{c}|^{2}/\partial\zeta^{2}$ and $\partial^{2}|\alpha_{c}|^{2}/\partial\delta^{2}$ with $\zeta$ and $\delta$ correspond to the dressed nearly-zero-energy crossings, and the extrema points of that correspond to the dressed nearly-zero-energy anticrossings. As shown in Fig. \ref{FIG:energy level zita-1}(a,d,g), the discontinuities of $\partial^{2}|\alpha_{c}|^{2}/\partial\zeta^{2}$ at $\zeta=-0.3$ in green and red lines align with the crossings, which correspond to the saltatorial intersections of $S$. The maxima of $\partial^{2}|\alpha_{c}|^{2}/\partial\zeta^{2}$ in green (at $\zeta=-0.21$) and cyan (at $\zeta=-0.02$) lines, and the minima of that in blue (at $\zeta=-0.21$) and green (at $\zeta=-0.02$) lines appear at the anticrossings, which correspond to the maxima of $S$. Similar patterns emerge throughout our spectral results. We also can see that the discontinuities in green (at $\zeta=0.07$), blue (at $\zeta=0.09$), and cyan (at $\zeta=0.22$) lines [Fig. \ref{FIG:energy level zita-1}(e)] match the crossings [Fig. \ref{FIG:energy level zita-1}(b)] at the saltatorial intersections of $S$ [Fig. \ref{FIG:energy level zita-1}(h)]. The maxima in green (at $\zeta=-0.03$) and magenta (at $\zeta=0.48$) lines, and the minima in magenta (at $\zeta=-0.03$) and cyan (at $\zeta=0.48$) lines [Fig. \ref{FIG:energy level zita-1}(e)] indicate the anticrossings [Fig. \ref{FIG:energy level zita-1}(b)] at the maxima of $S$ [Fig. \ref{FIG:energy level zita-1}(h)]. In Fig. \ref{FIG:energy level zita-1}(c,f,i), the maxima of $\partial^{2}|\alpha_{c}|^{2}/\partial\zeta^{2}$ in black (at $\zeta=-0.1$), green (at $\zeta=-0.05$), and blue (at $\zeta=0.03$) lines, and the minima of that in red (at $\zeta=-0.05$) and cyan (at $\zeta=0.03$) lines correspond to the anticrossings indicated by the maxima of $S$. Therefore, the second order partial derivatives of the photon Hopfield coefficients serve as robust indicators for identifying the spectral crossings and anticrossings. About $\partial^{2}|\alpha_{c}|^{2}/\partial\zeta^{2}$ in Fig. \ref{FIG:energy level zita-2} (d-f), and $\partial^{2}|\alpha_{c}|^{2}/\partial\delta^{2}$ in Fig. \ref{FIG:energy level delta-1} (e-i) and \ref{FIG:energy level delta-2} (e-i), the conclusion is similar as that of Fig. \ref{FIG:energy level zita-1}, thus it is no longer stated hereinbelow.

Comparing the spectra under different configurations, one can discover that the gap in the range $\zeta\in[-0.21,-0.02]$ for the configuration HO ST and its range [Fig. \ref{FIG:energy level zita-1}(a)] are respectively widened and extended by the heterogeneous coupling in multichains as shown in Fig. \ref{FIG:energy level zita-1}(b). For the mixed configuration HO-HE ST the new gaps near $\zeta=0$ appear as shown in Fig. \ref{FIG:energy level zita-1}(c). When the inter-chain coupling is TT, the spectra are shown in Fig. \ref{FIG:energy level zita-2}(a-c). It can be seen that the gaps near $\zeta=0$ narrow, and their ranges in $\zeta$ reduce compared with Fig. \ref{FIG:energy level zita-1}(a-c). In the space of $\delta$, the spectra are presented in Fig. \ref{FIG:energy level delta-1} (a-d) for even chain and in Fig. \ref{FIG:energy level delta-2} (a-d) for odd chain. In Fig. \ref{FIG:energy level delta-1} (a-d) we can see that the configuration HE ST has more crossings of the dressed nearly-zero-energy modes than HE TT. What's more, the crossings at $\delta=0.3$ and 0.36 [Fig. \ref{FIG:energy level delta-1} (c)] turn into anticrossing at $\delta=0.19$ [Fig. \ref{FIG:energy level delta-1} (d)] because of TT coupling. In the mixed multichains, the presence of the HO chain leads also to the decrease of the crossings, comparing Fig. \ref{FIG:energy level delta-1} (a) and \ref{FIG:energy level delta-1} (c). Comparing Fig. \ref{FIG:energy level delta-1} with Fig. \ref{FIG:energy level delta-2}, one can discover that the dressed nearly-zero-energy levels cross for even-length chain and avoid crossing for odd-length chain. Therefore, we can suppose that the coherent exchange is main mechanism affecting exciton transport for former, while the LZ transition is for latter.

In the next section, the steady-state dynamics with $\zeta$ and $\delta$ will be shown, in which the relation between the dressed nearly-zero-energy crossings (anticrossings) and the occupation number of photons are discussed. The exciton transport mechanism are revealed. The changes of exciton transport currents and efficiency with $\zeta$ and $\delta$ are explained. The optimal ranges of the inter-chain coupling $\zeta$, the dimerization parameter $\delta$, and the TLS transition frequency detuning $\Delta$ for exciton transport are found out by comparing the spectra and steady-state dynamics for different configurations.

\begin{figure*}[ht]
\centering
\includegraphics[width=17cm]{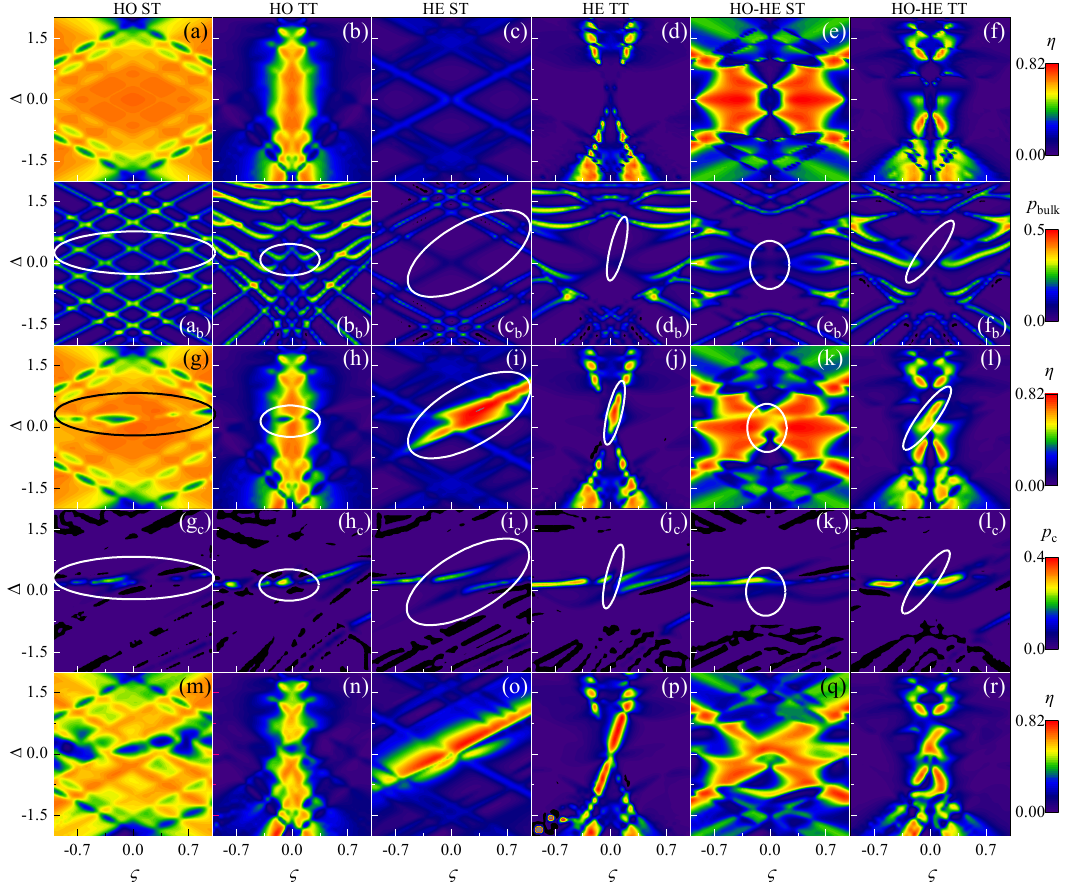}
\caption{\label{FIG:stady dynamic zita} (color online) The dependence of the steady-state dynamics on the inter-chain coupling strength $\zeta$ and the detuning $\Delta$ without cavity under TLS drive (a-f, a$_{b}$-f$_{b}$), and with cavity under TLS drive (g-l, g$_{c}$-l$_{c}$) and under TLS+cavity drive (m-r) for HO ST (the 1st column), HO TT (the 2ed column), $\cdots$, and HO-HE TT (the 6th column). The physical quantities of each row are written at right. The zones that the cavity induces significant effect on transport efficiency $\eta$ are marked by ellipses in (g-l), the corresponding zones are marked in (a$_{b}$-f$_{b}$) and (g$_{c}$-l$_{c}$). The parameters used are the same as those in Fig. \ref{FIG:stady dynamic zita withC line}.}
\end{figure*}
\begin{figure*}[ht]
\centering
\includegraphics[width=17cm]{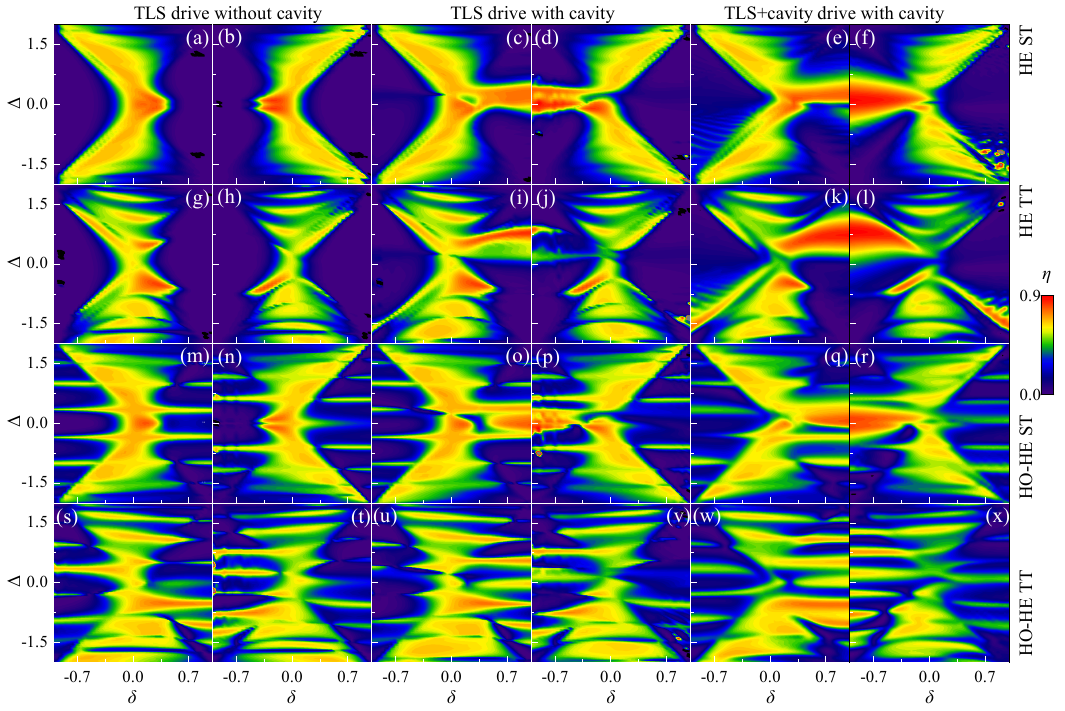}
\caption{\label{FIG:stady dynamic delta} (color online) The dependence of the steady-state transport efficiency $\eta$ on the dimerization parameter $\delta$ and the detuning $\Delta$, in which the conditions of the column and row are written at the top and right of this figure respectively. The parameters used are the same as those in Fig. \ref{FIG:stady dynamic zita withC line}.}
\end{figure*}
\begin{figure*}[ht]
\centering
\includegraphics[width=17cm]{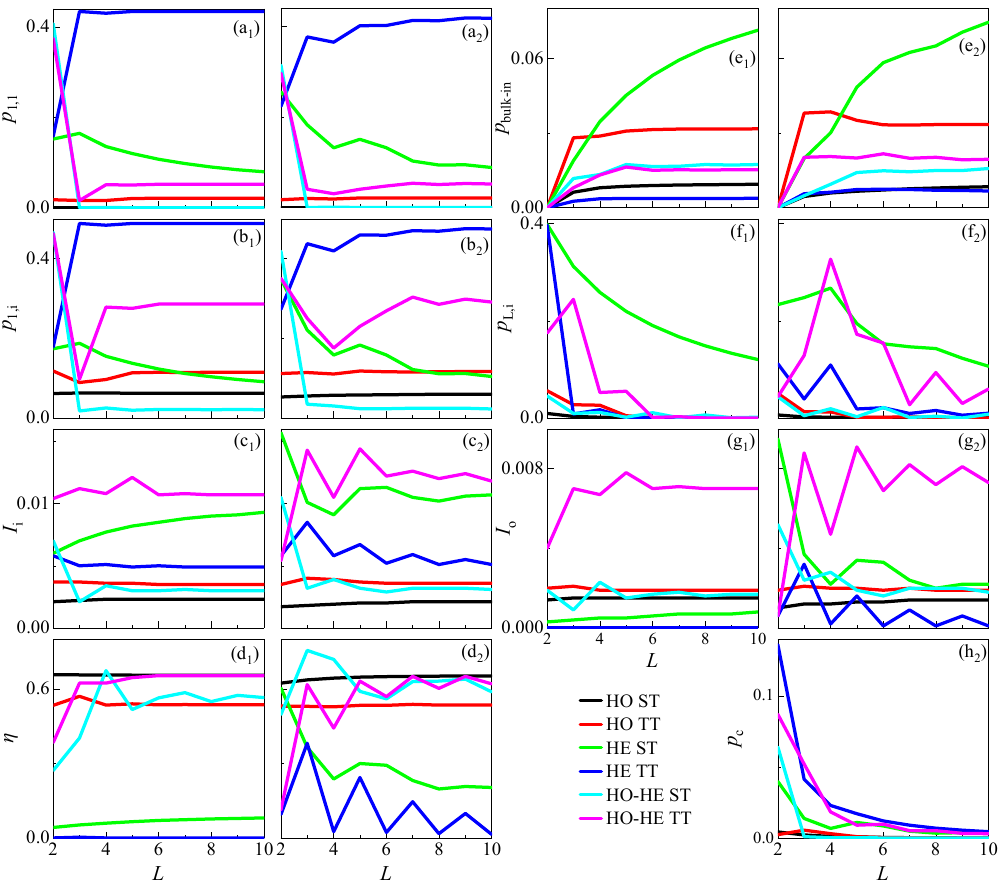}
\caption{\label{FIG:stady dynamic L} (color online) The steady-state dynamics as a function of the chain number $L$ without (a$_{1}$-g$_{1}$) and with (a$_{2}$-h$_{2}$) cavity for different configurations, in which $p_{1,i}$, $p_{L,i}$, and $p_{\rm bulk-in}$ represent respectively the occupation number of the first row, the last row, and the bulk state denoted as "bulk-in" containing all TLSs inside of the multichains. The parameter used is $N=10$, and the others are the same as those in Fig. \ref{FIG:stady dynamic zita withC line}.}
\end{figure*}
\begin{figure*}[ht]
\centering
\includegraphics[width=17cm]{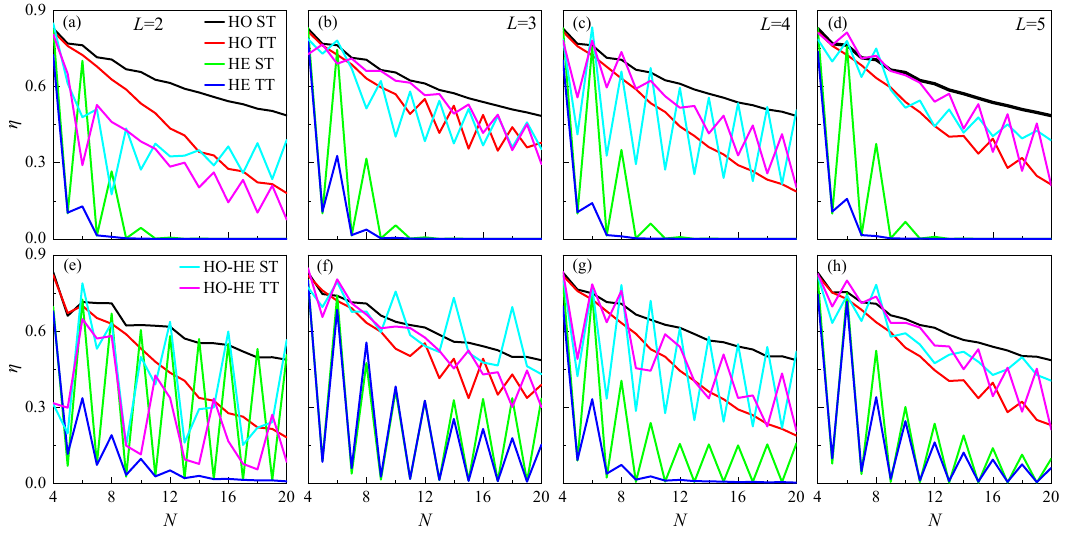}
\caption{\label{FIG:stady dynamic N 01} (color online) The steady-state dynamics as a function of the chain length $N$ for different configurations and chain number under TLS drive, (a-d) without cavity, (e-h) with cavity. The parameters used are the same as those in Fig. \ref{FIG:stady dynamic zita withC line}. }
\end{figure*}

\section{The exciton transport dynamics by multichains}\label{sec4}
In this section, we discuss the effects of the inter-chain coupling $\zeta$, the dimerization parameter $\delta$, the number $N$ and length $L$ of the chains on the exciton transport dynamics by numerically solving the master equation Eq. (\ref{eq:Lindblad master equation}) under an external driving, which are displayed in two subsections below. According to the analysis in the Sec. \ref{sec3}, the nearly-zero-energy modes are dressed by the cavity for strong exciton-cavity coupling, generating the polaritons. To ensure the polariton-dominated exciton transport, we set that the Rabi splitting ($\Omega_{R,k}$) is far stronger than both the cavity ($\kappa$) and TLSs ($\gamma$) decay rates, i.e., $\Omega_{R,k}\gg(\kappa,\gamma)$.

\subsection{The effects of the inter-chain coupling and the dimerization parameter}\label{subsec4.1}
Here, we investigate primarily the corresponding relationship between the cavity-dressed energy level crossings (anticrossings) and the exciton transport dynamics to elucidate the underlying transport mechanism. Based on the relation and the properties of exciton transport dynamics, we identify the optimal parameter ranges of the inter-chain coupling $\zeta$, the dimerization parameter $\delta$, and the TLS transition frequency detuning $\Delta$, as well as the optimal coupling configurations for enhanced exciton transport. Meanwhile, there are two physical quantities to be introduced, the steady-state occupation number $p_{11}$ of site (1,1) and $p_{\rm bulk}$ of bulk sites, where the site (1,1) denotes the first row and column in the TLS array shown in Fig. \ref{FIG:scheme graph01}, and the "bulk" site contains collective occupation of interior sites (columns 2 to N-1). This notation convention applies throughout unless specified otherwise. In addition, according to the Eq. (\ref{eq:outgoing exciton current 2}) the outgoing exciton current $I_{\rm o}$ is directly proportional to the total occupation number $\Sigma_{1}^{L}p_{k,N}$ at terminal sites. Hence, subsequent figures display only $I_{\rm o}$ as the transport metric, and $\Sigma_{1}^{L}p_{k,N}$ is denoted by $p_{N}$.

Fig. \ref{FIG:stady dynamic zita delta withoutC line} shows the dependence of the steady-state exciton transport dynamics of the multi-chain system without cavity on $\zeta$ and $\delta$ for the six chosen configurations. At first glance, one can discover that the inter-chain coupling and dimerization parameter exhibit distinct influences on exciton transport. The $p_{11}$, $p_{\rm bulk}$, $I_{\rm i,o}$, and $\eta$ with $\zeta$ are symmetric about $\zeta=0$ in $[-1,1]$, but with $\delta$ are not.

For the six configurations, the descending sort of efficiency is successively HO-HE ST, HO ST, HO TT, HO-HE TT, HE ST, HE TT in $\zeta\in[-1,1]$ [Fig. \ref{FIG:stady dynamic zita delta withoutC line}(a$_{5}$)]. For the HO-HE ST, near $\zeta=0.3$ the sites of bulk [Fig. \ref{FIG:stady dynamic zita delta withoutC line}(a$_{2}$)], input, and outgoing possess simultaneously relatively high occupation numbers of exciton, which support exciton transport in multichains. Consequently, this behaves the largest exciton transport efficiency, input and outgoing currents [Fig. \ref{FIG:stady dynamic zita delta withoutC line}(a$_{3, 4, 5}$)]. On the contrary, at $\zeta=0$ and $\zeta\rightarrow\pm1$, despite more excitations at the site (1,1) [Fig. \ref{FIG:stady dynamic zita delta withoutC line}(a$_{1}$)], the exciton occupation number $p_{N}$ at outgoing sites is too few to transfer to the drain, leading to the low efficiency and exciton currents. Specially, the excitons are distributed mainly at bulk sites for $\zeta\rightarrow\pm1$ [see cyan line in Fig. \ref{FIG:stady dynamic zita delta withoutC line}(a$_{2}$)], which is beneficial to energy storage and has potential applications for designing quantum battery \cite{quach2022superabsorption}. Except for the HE ST, the other configurations also have the property for $\zeta\rightarrow\pm1$, as well as HO TT and HO-HE TT do near $\zeta=\pm0.4$. However, for HO ST near $\zeta=\pm0.35$ and $\pm0.97$, $p_{\rm bulk}$, $I_{\rm i,o}$, and $\eta$ are relatively high, thus the abilities both of energy transport and storage are strong. Near $\zeta=\pm0.7$ for HO ST, although the exciton transport efficiency is large, exciton currents are relatively small. Thus it is not beneficial to the exciton transport.

The system has four configurations HE ST, HE TT, HO-HE ST, HO-HE TT containing HE chain. Because of the different spectra between odd- and even-length HE chain, Fig. \ref{FIG:stady dynamic zita delta withoutC line}(b$_{1-5}$) compares the steady-state dynamics with $\delta$ for $N=8,~9$ in the absence of cavity. It is obvious that the exciton currents and transport efficiency are the largest near $\delta=0.3$ for the configurations HE ST and HO-HE ST of $N=8$ than that in the other range of $\delta$ and for the other configurations [Fig. \ref{FIG:stady dynamic zita delta withoutC line}(b$_{3-5}$)], which are good for effective exciton transport. However, the HE TT of $N=9$ is the most beneficial to energy storage because of the high $p_{\rm bulk}$ [Fig. \ref{FIG:stady dynamic zita delta withoutC line}(b$_{2}$)] as well as the lowest $I_{\rm i,o}$ and $\eta$. Particularly, the exciton transport is anomalous in $\delta<0$, where the efficiency is high, but the currents approach to zero, as shown in Fig. \ref{FIG:stady dynamic zita delta withoutC line}(b$_{3-5}$), thus there is no real exciton transport. The reason is that the drive fails to coherently excite the site (1,1), namely the excitons can not be injected in multichains [Fig. \ref{FIG:stady dynamic zita delta withoutC line}(b$_{1,2}$)]. When $\delta>0.35$, the exciton currents and transport efficiency decrease sharply with $\delta$ or too small, because the excitons are localized at the left-edge site so that the occupation number of the right-edge sites is small for the HO-HE ST and HO-HE TT of $N=8,9$ or approaches to zero for the others. 

Considering a cavity, we plot the Fig. \ref{FIG:stady dynamic zita withC line} and \ref{FIG:stady dynamic delta withC line} corresponding to Fig. \ref{FIG:stady dynamic zita delta withoutC line}, which present the effects of the inter-chain coupling $\zeta$ and the dimerization parameter $\delta$ on the exciton transport dynamics. Our analysis specifically focuses on the change of the steady-state dynamics with $\zeta$ and $\delta$ at the cavity-dressed nearly-zero-energy crossings and anticrossings. In the multichain-cavity coupled system, the main channels of exciton transport contain the multichains and polaritons which depend on occupation number of excitons and photons, while it only is the multichains in the absence of the cavity.

Under the TLS and the TLS+cavity drive, the effect of the inter-chain coupling $\zeta$ on the steady-state dynamics of the multi-chain system in a cavity is presented for different configurations in Fig. \ref{FIG:stady dynamic zita withC line}(a$_{1-6}$) and \ref{FIG:stady dynamic zita withC line}(b$_{1-6}$), respectively. Particularly, to analyse what role the cavity acts as in exciton transport, the dependence of the photon occupation number $p_{\rm c}$ on $\zeta$ is shown in Fig. \ref{FIG:stady dynamic zita withC line}(a$_{5}$) and \ref{FIG:stady dynamic zita withC line}(b$_{5}$), in which the extreme points of $p_{\rm c}$ are marked by the vertical fine lines. It can be found through comparison that the extreme points correspond to the dressed nearly-zero-energy crossings and anticrossings in spectra, as well as the discontinuity and extreme points of the second order partial derivative $\partial^{2}|\alpha_{c}|^{2}/\partial\zeta^{2}$ of the photon Hopfield coefficients marked by the vertical fine lines in Fig. \ref{FIG:energy level zita-1} and \ref{FIG:energy level zita-2}. In consequence, $\partial^{2}|\alpha_{c}|^{2}/\partial\zeta^{2}$ can track the change of $p_{\rm c}$ with $\zeta$. Comparing with Fig. \ref{FIG:stady dynamic zita delta withoutC line}, one can discover that the presence of the cavity breaks the symmetry of $I_{\rm i,o}$ with respect to $\zeta=0$ for configurations HO ST, HE ST, HE TT, and HO-HE ST [Fig. \ref{FIG:stady dynamic zita withC line}(a$_{3,4}$)]. Moreover, the exciton transport is suppressed by the cavity in the range of $\zeta\neq0$. For example, $p_{\rm c}$ is high near $\zeta=-0.3$ [Fig. \ref{FIG:stady dynamic zita withC line}(a$_{5}$)], suppressing the exciton currents in the main channel with the multichains for HO ST [Fig. \ref{FIG:stady dynamic zita withC line}(a$_{3,4}$)]. On the other hand, polariton-assisted transport boosts $I_{\rm i,o}$ and $\eta$ near $\zeta=0$ for configurations HE ST, HE TT, and HO-HE TT, in which $p_{\rm c}$ is high, leading the increase of exciton occupation number $p_{N}$ on the last column sites [Fig. \ref{FIG:stady dynamic zita withC line}(a$_{4}$)]. In fact, $\zeta=0$ means that only one chain undertakes the exciton transport, in which polariton-assisted exciton transport is demonstrated \cite{wei2022cavity}. Because of few excitations injected [Fig. \ref{FIG:stady dynamic zita withC line}(a$_{1}$)] for HO ST, HO TT and few $p_{N}$ [Fig. \ref{FIG:stady dynamic zita withC line}(a$_{4}$)] for HO-HE ST near $\zeta=0$, the cavity has no effect on exciton transport. What's more, it can be seen obviously that the peaks of $p_{\rm c}$ and of $I_{\rm i,o}$ are at different locations. The reason is that at the peak of $p_{\rm c}$ the transfer of excitation in the multichains to the cavity suppresses the exciton transport that the multichains server as dominating channel, then causing relatively low $I_{\rm i,o}$. Adding the cavity drive, $p_{\rm c}$ increase dramatically but without proportional $I_{\rm i,o}$ and $\eta$ enhancement as shown in Fig. \ref{FIG:stady dynamic zita withC line}(b$_{5}$). Because of the transfer of excitations, $p_{11}$ increases near $\zeta=0$ [Fig. \ref{FIG:stady dynamic zita withC line}(b$_{1}$)], causing $p_{\rm c}$ decrease for HE ST, HE TT, HO-HE ST, and HO-HE TT [Fig. \ref{FIG:stady dynamic zita withC line}(b$_{5}$)], and suppressing the exciton transport of the main channel with the porlaritons [Fig. \ref{FIG:stady dynamic zita withC line}(b$_{3,4,6}$)]. However, one can see by comparing Fig. \ref{FIG:stady dynamic zita withC line}(b$_{3,4}$) and Fig. \ref{FIG:stady dynamic zita withC line}(a$_{3,4}$) that the increase of $p_{\rm c}$ suppresses significantly the exciton currents of the main channel with the multichains near $\zeta=0.2$ for HO-HE ST and near $\zeta=0.35$ for HO ST. Therefore, the exciton transport can be adjusted by the cavity photon. For different configurations, we have the conclusion that ST coupling outperforms TT for the exciton transport. Besides, the optimal $\zeta$ ranges differ for HO and HE chains, which are consistent with the spectral features. In a word, the photon-exciton competition governs the suppression and enhancement of exciton transport via cavity. 

Fig. \ref{FIG:stady dynamic delta withC line}(a$_{1-6}$) and \ref{FIG:stady dynamic delta withC line}(b$_{1-6}$) show the dependence of the steady-state dynamics on the dimerization parameter $\delta$ under the TLS and the TLS+cavity drive in the presence of cavity, respectively. We analyse firstly the effects of $\delta$ and cavity on exciton transport under the TLS drive. Although the exciton transport efficiencies are centered approximatively on $\delta=0$ [Fig. \ref{FIG:stady dynamic delta withC line}(a$_{6}$)], there are no the exciton currents in $\delta<0$ except for the configurations HE ST and HO-HE ST of $N=9$ [Fig. \ref{FIG:stady dynamic delta withC line}(a$_{3,4}$)]. The root cause is the blocked excitation injection at the site (1,1) [Fig. \ref{FIG:stady dynamic delta withC line}(a$_{1,3}$)]. It can be seen that for the configurations HE ST and HO-HE ST of $N=8$ in the range $0<\delta<0.35$ the presence of cavity suppresses significantly the exciton currents $I_{\rm i,o}$ [Fig. \ref{FIG:stady dynamic delta withC line}(a$_{3,4}$)] and transport efficiency $\eta$ [Fig. \ref{FIG:stady dynamic delta withC line} (a$_{6}$)] by comparing with Fig. \ref{FIG:stady dynamic zita delta withoutC line}(b$_{3-5}$). Particularly, at $\delta=0.26$ for HE ST and $\delta=0.36$ for HO-HE ST of $N=8$, the troughs of $I_{\rm i,o}$ and $\eta$ appear, which correspond to the peak and the trough of $p_{\rm c}$ [Fig. \ref{FIG:stady dynamic delta withC line}(a$_{5}$)] and to the dressed
nearly-zero-energy crossings [Fig. \ref{FIG:energy level delta-1}], respectively. The reason for the former is that the excitations on the edge-column TLSs are transferred to cavity through coherent exciting at the crossings as shown in Fig. \ref{FIG:stady dynamic delta withC line}(a$_{1,3,4}$), suppressing the exciton currents and transport efficiency on the main channel with multichains. But for the latter, the occupation number $p_{N}$ is too few, limiting the exciton transport. At $\delta>0.26$ for HE ST and $\delta>0.36$ for HO-HE ST of $N=8$, the limited photon number $p_{\rm c}$ [Fig. \ref{FIG:stady dynamic delta withC line}(a$_{5}$)] and the decreased occupation number $p_{\rm bulk}$ [Fig. \ref{FIG:stady dynamic delta withC line} (a$_{2}$)] lead to that the polariton becomes the main channel, thus the exciton transport is strengthened by the cavity comparing with Fig. \ref{FIG:stady dynamic zita delta withoutC line}(b$_{3,4}$). However, for even-length HO-HE TT both the multichains and the polariton dominate the exciton transport because of the large $p_{\rm bulk}$ and $p_{\rm c}$. For odd-length HO-HE TT, in $\delta\gtrsim0.35$ the low $p_{N}$ is main reason limiting exciton transport. When the cavity drive is added based on the TLS drive, there are distinct effects on exciton transport in $0<\delta<0.35$ for HE ST and HO-HE ST of $N=8$. The increased $p_{\rm c}$ [Fig. \ref{FIG:stady dynamic delta withC line}(b$_{5}$)] leads to more polaritons, enhancing the polariton channel [Fig. \ref{FIG:stady dynamic delta withC line}(b$_{3,4}$)] than that under TLS drive [Fig. \ref{FIG:stady dynamic delta withC line}(a$_{3,4}$)], but in $\delta>0.35$ because of the low $p_{\rm c}$ and $p_{\rm bulk}$ [Fig. \ref{FIG:stady dynamic delta withC line}(b$_{2}$)] the exciton transport is limited for even-length HE ST. On the contrary, because of the large $p_{\rm c}$ and $p_{\rm bulk}$ the exciton transport is heightened for HO-HE ST and HO-HE TT with odd length, as well as for HO-HE ST in $\delta>0.7$ compared with that under TLS drive. Particularly, the exciton transport currents and efficiency are significantly large due to the cavity drive in $\delta<0$ for all configurations of $N=9$, which break the conventional preference that the transport are superior in $\delta>0$ than that in $\delta<0$. In the condition, the high $p_{\rm bulk}$ [Fig. \ref{FIG:stady dynamic delta withC line}(b$_{2}$)] and $p_{\rm c}$ [Fig. \ref{FIG:stady dynamic delta withC line}(b$_{5}$)] demonstrate that both the multichains and the polaritons participate the exciton transport. Interestingly, in the presence of the cavity drive $p_{11}$ is no longer necessary to transport efficiently exciton [Fig. \ref{FIG:stady dynamic delta withC line}(b$_{1}$)], but $p_{N}$ still is. For instance, the low $p_{N}$ limits the exciton transport for HE TT and HO-HE TT of $N=9$ in $\delta<0$. To sum up, in $\delta>0$ the cavity have significant influence on exciton transport for even-length multichains under TLS drive, which is suppressed in weak $\delta$ and is enhanced in strong $\delta$. In the space of $\delta$, even-length ST is more beneficial to exciton transport than TT under TLS drive, but odd-length ST is under TLS+cavity drive. In addition, no matter which one is main for multichain and polariton channels, the occupation of terminal sites decides the exciton transport.

When the TLS transition frequency detuning $\Delta$ is considered, we plot the steady-state dynamics of the hybrid system in Fig. \ref{FIG:stady dynamic zita} and \ref{FIG:stady dynamic delta}, which show respectively the effects of the inter-chain coupling $\zeta$ and the dimerization parameter $\delta$ for different configurations. Comparing with Fig. \ref{FIG:energy level zita-1} - \ref{FIG:energy level delta-2}, the effects of $\zeta$, $\delta$, and $\Delta$ on $\eta$ mirror spectral structures. 

In Fig. \ref{FIG:stady dynamic zita} the zones of $\zeta$ and $\Delta$ that the cavity has obviously effects on exciton transport are marked in Fig. \ref{FIG:stady dynamic zita}(a$_{b}$-f$_{b}$, g-l, and g$_{c}$-l$_{c}$) by ellipses through comparing to Fig. \ref{FIG:stady dynamic zita}(a-f). According to the effects these coupling configurations of the multichains can be classified into two categories. One is HO ST and HO TT that the exciton transport is suppressed, the others has opposite influence. For the former, the excitation on whole multichains transfer to the cavity disrupts the multichains transport serving as dominating channel [Fig. \ref{FIG:stady dynamic zita}(g$_{c}$,h$_{c}$)]. For the letter, as shown in Fig. \ref{FIG:stady dynamic zita}(c$_{b}$-f$_{b}$) too small $p_{\rm bulk}$ limits the multichain channel in the marked zone, but the large $p_{\rm c}$ strengthens the polariton channel [Fig. \ref{FIG:stady dynamic zita}(i$_{c}$-l$_{c}$)]. We can further illustrate the mechanism through the spectra. Interestingly, the structures of $p_{\rm c}$ [Fig. \ref{FIG:stady dynamic zita}(g$_{c}$-l$_{c}$)] are similar to those of the cavity-dressed energy levels [Fig. \ref{FIG:energy level zita-1} and \ref{FIG:energy level zita-2}]. In the intensified zone by the cavity the cavity-dressed anticrossings exist [Fig. \ref{FIG:energy level zita-1} and \ref{FIG:energy level zita-2}], thus LZ transition leads to that the excitations on multichains generate polaritons. Particularly, in the zone with high efficiency there are two anti-crossed zonate zones with large $p_{\rm c}$ [Fig. \ref{FIG:stady dynamic zita}(i$_{c}$)] corresponding to the cavity-dressed anticrossings for HE ST. When the cavity drive is exerted, for the two categories of configurations the exciton transport is further suppressed and enhanced by the cavity respectively. HE ST shows the biggest effect of cavity drive on exciton transport in six configurations, and that the high efficient zone is significantly extended. 

In parameter space of $\delta$ and $\Delta$, as shown in Fig. \ref{FIG:stady dynamic delta} the steady-state dynamics depend on the coupling configuration, the cavity, and the method of drive. For HE ST, HE TT, and HO-HE ST of $N=8,~9$ the cavity has apparent effects on exciton transport efficiency in the zone of the fully dimerized limit, namely, $\delta\rightarrow\pm1$, and near the zero detuning [Fig. \ref{FIG:stady dynamic delta}(a-d, g-i, m-p)]. Nevertheless, it is contrary for HO-HE TT [Fig. \ref{FIG:stady dynamic delta}(s-v)] because the HO chain and the diagonal inter-chain coupling suppress the dimerized effect. Undoubtedly, in the zone enhancing exciton transport the polaritons act as prime role whose mechanism is coherent excitation (LZ transition) located at the cavity-dressed crossings (anticrossings) according to the analysis on Fig. \ref{FIG:energy level delta-1}, \ref{FIG:energy level delta-2} and \ref{FIG:stady dynamic delta withC line}. When the cavity drive is added, it can be seen that the high efficient zones are extended and the exciton transport efficiency is strengthened in the zone of $\delta\rightarrow\pm1$ and near $\Delta=0$ for HE ST, HE TT, and HO-HE ST of $N=8,~9$ [Fig. \ref{FIG:stady dynamic delta}(e, f, k, l, q, r)]. In point of the optimal exciton transport, HE ST is the preferred configuration in the parameter space of $\zeta$ and $\Delta$ under the TLS drive in the presence of the cavity, as well as HE TT is the preferred configuration in the parameter space of $\delta$ and $\Delta$ under the TLS+cavity drive. Significantly, according to the results the exciton transport can be controlled by the coupling configurations, the cavity, the drive methods, the inter-chain coupling, the dimerization parameter, and the TLS transition frequency detuning.
\subsection{The effects of the number and the length of chains}\label{subsec4.2}
In this part, we discuss the dependence of exciton transport dynamics on the chain number $L$ and length $N$, which reveal the inter-chain joint effect, and give optimal $L$, $N$, and configurations with high transport efficiency and currents in Fig. \ref{FIG:stady dynamic L} and \ref{FIG:stady dynamic N 01}. 

Fig. \ref{FIG:stady dynamic L}(c$1$, g$1$, d$1$) shows that in the absence of the cavity the exciton transport is mightily dependent on chain number $L$ for the three configurations HE ST, HO-HE ST, and HO-HE TT when $L\leq6$, and is opposite for the others. For HE TT, although the input exciton current $I_{i}$ is large, the little $p_{N}$ limits the outgoing exciton current $I_{o}$, leading to that exciton transport is independent of $L$. In fact, the inefficient exciton transport among chains lead to excitons localizing at the upper edge sites [Fig. \ref{FIG:stady dynamic L}(b$_{1}$)], which is similar to skin effect. Contrary to the skin effect, for HE ST because of the diffusion of excitons among chains the inner occupation number $p_{\rm bulk-in}$ increase and that of the edge decrease with $L$ as shown in Fig. \ref{FIG:stady dynamic L}(b$_{1}$, e$_{1}$, f$_{1}$). For large $L$ the excitons collect at the inside of the multichains [Fig. \ref{FIG:stady dynamic L}(e$_{1}$)], thus HE ST is a potential candidate applied to quantum battery in the six configurations. In the presence of the cavity the steady-state dynamics present characteristic phenomena as shown in Fig. \ref{FIG:stady dynamic L}(a$_{2}$-h$_{2}$). The prominent one is that the exciton currents and transport efficiency present distinctly odd-even oscillation with $L$ for HE TT and HO-HE TT, in which the peaks (troughs) of $I_{\rm i,o}$ and $\eta$ correspond to the odd (even) $L$ when $N=10$ [Fig. \ref{FIG:stady dynamic L}(c$_{2}$, g$_{2}$, d$_{2}$)]. The difference between the steady-state dynamics of the odd and even $L$ becomes small with $L$. In addition, it can be seen that only for HE ST and HE TT the exciton transport is strengthened by the cavity when $N=10$.

There is an obvious character in Fig. \ref{FIG:stady dynamic N 01} that the exciton transport efficiency $\eta$ also behaves odd-even oscillation with the chain length $N$. When the cavity is absent, the oscillation is weak for HO ST relative to the other configurations and depends simultaneously on the odevity of $L$ and $N$ for HO TT [Fig. \ref{FIG:stady dynamic N 01}(a-d)]. Thus the diagonal inter-chain coupling induces the difference between the exciton transports of odd and even $N$ in long HO TT multichains as shown in Fig. \ref{FIG:stady dynamic N 01}(b, d). It can be found that because of the dimerization the prominent oscillation with $N$ exists in the short multichains, but disappears in the long multichains for HE ST and HE TT [Fig. \ref{FIG:stady dynamic N 01}(a-d)]. In fact, the dimerization hinders the exciton diffusing on the multichains for the later. However, in the mixed long multichains of HO-HE, because HO chains suppress the dimerization, the oscillation is enhanced for HO-HE ST and HO-HE TT. For the two configurations the oscillations with $N$ are consistent for odd $L$ and are opposite for even $L$ each other, such as when $L=3$ and $N=16$ $\eta$ is trough for both HO-HE TT and HO-HE ST, while $L=4$ and $N=16$ $\eta$ is peak for HO-HE ST and is trough for HO-HE TT. When the cavity is considered the oscillation is significantly strengthened for the configurations containing HE chain, as shown in Fig. \ref{FIG:stady dynamic N 01}(e-h). For HE ST the oscillation decreases with $L$, which is caused by the dissipation of the more and more TLSs. Whatever situation the oscillation of $\eta$ with $N$ for HE ST and HE TT is consistent, which illustrates further that the oscillation is mainly caused by HE chain. Interestingly, the oscillation amplitude behaves inverse dependence on the odevity of $L$ between HE TT and HO-HE ST, HO-HE TT. Concretely, for the former the amplitude with odd $L$ is bigger than that with even $L$, and for the latter it is contrary. Furthermore, it can be discovered that for HO-HE ST and HO-HE TT the odd-even oscillation of $\eta$ with $N$ can be adjusted by $L$ and the cavity. For example, the oscillations of $L=3$ and $L=4$ for HO-HE ST are out of phase [Fig. \ref{FIG:stady dynamic N 01}(a, b)]. $\eta$ without [Fig. \ref{FIG:stady dynamic N 01}(b)] and with [Fig. \ref{FIG:stady dynamic N 01}(f)] cavity for HO-HE ST of $N=14$ and $L=2$ are respectively trough and peak. Comparing Fig. \ref{FIG:stady dynamic N 01}(b) and \ref{FIG:stady dynamic N 01}(f), Fig. \ref{FIG:stady dynamic N 01}(f) and \ref{FIG:stady dynamic N 01}(g) for HO-HE ST, the period of the oscillation also is varied by the cavity and $L$.
\section{Summary}\label{sec5}
In summary, in order to reveal the exciton transport mechanism of a 2D multichains system within a cavity, we investigate systematically the spectra through numerically diagonalizing the Hamiltonian, the second order partial derivative of photon Hopfield coefficients with respect to the inter-chain coupling strength and the dimerization parameter, the von Neumann entropy and the steady-state dynamics of exciton via solving the Lindblad quantum master equation, in which six inter-TLS coupling configurations are considered as HO ST, HO TT, HE ST, HE TT, HO-HE ST and HO-HE TT based on the HO and HE chain, as well as the square and triangular inter-chain coupling type. The results show that the exciton transport can be adjusted by the inter-chain coupling strength, the dimerization parameter, the detuning of the driving frequency, the chain number and length, the coupling configurations, and the cavity. In the absence of cavity the exciton transport is decided by the distribution of exciton on whole multichains, in which the efficient transport need the excitons spread all over the multichains. Further, when the cavity is considered it is governed by the competition between the polariton and multichain channel. When the excitations are mainly distributed on whole multichains, the multichains dominate the transport. Otherwise, the cavity occupies the most excitations, the polaritons act as the main channel, meanwhile the limited exciton occupations of the multichain terminals are necessary. The exchange condition of the excitation between the multichains and cavity is the key which channel is primary. The number of polaritons depend on the photon occupation number, which is relate to the cavity-dressed nearly-zero-energy crossings and anticrossings. At the crossings and anticrossings the coherent excitation and LZ transition between the multichains and cavity occur, respectively. Therefore, the exciton transport is controlled by the crossings and anticrossings, whose locations can be accurately determined by von Neumann entropy. Besides, we discover that the second order partial derivative of the photon Hopfield coefficient serves as robust indicator for identifying the crossings and anticrossings, and the change of photon occupation number. The spectra show that the crossings and anticrossings are related closely to the coupling configurations, the inter-chain coupling strength, and the dimerization parameter. In the space of inter-chain coupling strength, ST outperforms TT for the exciton transport, but the optimal ranges differ for HO and HE chains. In the space of the dimerization parameter, even-length ST is more beneficial to exciton transport than TT under TLS drive, but odd-length ST is under TLS+cavity drive. Interestingly, when the cavity drive is exerted, the exciton transport is not limited by the occupation number of site (1,1) and is strengthened for HE ST and HO-HE ST with odd chains in the range of the negative dimerization parameter. Further, we discover that the steady-state dynamics can be adjusted by the TLS transition frequency detuning. Finally, it can be found that the exciton transport currents and efficiency present distinctly odd-even oscillation with chain length and number, which depends on the coupling configuration and the cavity. In a word, our results reveal the exciton transport mechanism of the 2D hybrid system, which provides theoretical foundation for designing controllable and high efficient exciton transport devices, such as solar cell. Of course, it has many points of view to reveal the mechanism, like the topological phase transition of the system.
 
\section{Acknowledgements}\label{sec6}
This work is supported by the Academic Ability Promotion Foundation for Young Scholars of Northwest Normal University in China under the Grant No. NWNU-LKQN2024-10, National Natural Science Foundation of China under Grant No. 12164042, National Natural Science Foundation of China under Grant No. 12104374, National Natural Science Foundation of China under Grant No. 12065022.

\end{document}